\documentclass[a4paper,11pt]{article}
\usepackage{jheppub} 
\usepackage{mathtools}
\usepackage{slashed}

\newcommand{\phiend}{\phi_{\text{end}}}
\newcommand{\Trh}{T_\text{rh}}
\newcommand{\Tmax}{T_\text{max}}
\newcommand{\gs}{g_\star}
\newcommand{\gss}{g_{\star s}}

\newcommand{\arh}{a_\text{rh}}
\newcommand{\aend}{a_\text{end}}
\newcommand{\amax}{a_\text{max}}
\newcommand{\rp}{\rho_\phi}
\newcommand{\rR}{\rho_R}
\newcommand{\mdm}{m_\chi}

\title{\Large Graviton- and Inflaton-mediated Dark Matter Production\\after Large Field Polynomial Inflation}
\author[a]{Nicolás Bernal,}
\author[b]{Julia Harz,}
\author[b,c]{Martin A. Mojahed}
\author[b]{and Yong Xu}
\affiliation[a]{New York University Abu Dhabi\\
	PO Box 129188, Saadiyat Island, Abu Dhabi, United Arab Emirates}
\affiliation[b]{\it PRISMA$^+$ Cluster of Excellence and Mainz Institute for Theoretical Physics\\
	Johannes Gutenberg University, 55099 Mainz, Germany}
\affiliation[c]{Physik Department, TUM School of Natural Sciences, TU M\"unchen,\\ James-Franck-Stra{\ss}e, D-85748 Garching, Germany}

\emailAdd{nicolas.bernal@nyu.edu}
\emailAdd{julia.harz@uni-mainz.de}
\emailAdd{mojahedm@uni-mainz.de}
\emailAdd{yonxu@uni-mainz.de}

\abstract{Polynomial inflation is a simple cosmological scenario, which fits the cosmic microwave background data well. It provides testable predictions for the tensor-to-scalar ratio and the running of the spectral index. In this work, we investigate the production of Dirac dark matter (DM) within the framework of large-field polynomial inflation. We study all relevant production channels including $i$) non-thermal production through inflaton decays and scatterings, and $ii$) thermal production from scattering of standard model particles mediated by inflatons and gravitons. In contrast to small-field polynomial inflation, where inflaton decay dominates DM production, we find that graviton-mediated processes can be dominant in the large-field scenario. For DM lighter than the inflaton, we demonstrate that the interplay between graviton- and inflaton-mediated production channels give rise to non-trivial relations between the DM mass and the reheating temperature required to account for the DM relic abundance.}

\begin{document} 
	\begin{flushright}
		MITP-24-056
	\end{flushright}
	\maketitle
	\flushbottom
	
	%%%%%%%%%%%%%%%%%%%%%%%%%%%%%%%%%%%%%%%%%%%%%
	\section{Introduction}
	%%%%%%%%%%%%%%%%%%%%%%%%%%%%%%%%%%%%%%%%%%%%%
	Cosmic inflation is an established paradigm for solving the horizon and flatness problem of standard cosmology~\cite{Starobinsky:1980te, Guth:1980zm, Linde:1981mu, Albrecht:1982wi}. The simplest inflationary scenario is the so-called slow roll (SR) single-field inflation, where a scalar inflaton field $\phi$ slowly rolls down its potential~\cite{Martin:2013tda}. For the minimal SR model with a monomial potential $V(\phi) \propto \phi^p$, the most recent cosmic microwave background (CMB) experiments~\cite{Planck:2018jri, BICEP2:2018kqh} have ruled out scenarios with $p \geq 1/2$, since such potentials are too steep and yield too large tensor-to-scalar ratios $r$. Moreover, according to the latest Planck data~\cite{Planck:2018jri}, the most favored single-field inflation models within the Einstein gravity framework are those with concave potentials $V^{\prime \prime}(\phi)<0$. 
	
	As a simple extension of the monomial paradigm, one can consider a renormalizable polynomial potential with a  concave shape that contains a quadratic, cubic, and a quartic term~\cite{Hodges:1989dw, Allahverdi:2006iq, Destri:2007pv, Nakayama:2013jka, Nakayama:2013txa, Kallosh:2014xwa, Li:2014zfa, Aslanyan:2015hmi, Gao:2015yha, Musoke:2017frr}. The cosmic inflationary dynamics of this model, as well as the parameter space that is in agreement with CMB data, have recently been studied in the context of small- and large-field excursions (Refs.~\cite{Drees:2021wgd} and~\cite{Drees:2022aea}, respectively). After the end of inflation, the energy density contained in the inflaton field is transferred to standard-model (SM) degrees of freedom producing the SM thermal plasma, for instance via inflaton decays. 
	
	It is well known that physics beyond the standard model (BSM) is required to explain both the existence of dark matter (DM) and the dynamical generation of the baryon asymmetry of the Universe. In the context of polynomial inflation, a rather large parameter space was found for both thermal and non-thermal baryogenesis via leptogenesis~\cite{Xu:2022qpx, Drees:2024hok}. DM can be produced by several processes, including non-thermal production from inflaton decays and scatterings, and thermal production from inflaton- and graviton-mediated scatterings of SM particles in the thermal plasma. In the context of small-field polynomial inflation, the generation of fermionic DM was investigated in Ref.~\cite{Bernal:2021qrl} where it was shown that inflaton decays dominate over all other production channels. 
	
	In this work, we extend the previous work~\cite{Bernal:2021qrl} by considering a large-field setup~\cite{Drees:2022aea}. 
	In contrast to the previously studied small-field case, we find that DM production channels mediated by gravity or the inflaton can become important. Here we present a detailed account of the interplay of DM production via gravity-mediated processes and inflation decays and scatterings. Although the current analysis is focused on polynomial inflation, our formalism can be applied to other large-field inflation models such as Starobinsky inflation~\cite{Starobinsky:1980te}, Higgs inflation~\cite{Bezrukov:2007ep} or attractor models~\cite{Kallosh:2013yoa}.
	
	The outline of this paper is as follows. We introduce the model setup in Section~\ref{sec:setup}. The dynamics of inflation and (p)reheating are discussed in Section~\ref{sec:inflation_reheating}. It will set the stage for the main focus of this work, the production of DM, which is investigated in Section~\ref{sec:dm}. Here, we analyze all possible production channels and the corresponding parameter space to successfully explain the DM abundance. We summarize our findings in Section~\ref{sec:conclusion}.
	
	%%%%%%%%%%%%%%%%%%%%%%%%%%%%%%%%%%%%%%%%%%%%%%
	\section{Model Setup} \label{sec:setup}
	%%%%%%%%%%%%%%%%%%%%%%%%%%%%%%%%%%%%%%%%%%%%%%
	Inspired by Occam's razor, and the necessity of new physics explaining the primordial evolution of the Universe and DM, we extend the SM of particle physics with two new particles, which account for the inflaton $\phi$, and DM $\chi$. The action $S$ of the model can be written as follows~\cite{Bernal:2021qrl}
	\begin{equation} \label{eq:action}
		S = \int d^4x\, \sqrt{-g} \left(\mathcal{L}_{\rm EH} +  \mathcal{L}_{\rm{SM}} + \mathcal{L}_{\phi}+ \mathcal{L}_{\text{DM}} \right),
	\end{equation}
	where $g$ is the determinant of the Friedmann-Lemaître-Robertson-Walker metric defined by $g_{\mu \nu} = \text{diag}(1,-a^2,-a^2,-a^2)$, and $a$ is the cosmic scale factor. $\mathcal{L}_{\rm EH}$ denotes the Einstein-Hilbert Lagrangian density given by
	\begin{equation}
		\mathcal{L}_{\rm EH}  = \frac{M_P^2}{2}\, R\,,
	\end{equation}
	where $R$ denotes the Ricci scalar and $M_P \simeq 2.4 \times 10^{18}$~GeV is the reduced Planck mass. $\mathcal{L}_{\rm SM}$ is the usual SM Lagrangian and $\mathcal{L}_{\phi}$ encodes the inflaton dynamics
	\begin{equation}
		\mathcal{L}_{\phi} = \frac{1}{2} \partial_{\mu} \phi \partial^{\mu} \phi - V(\phi)\,.
	\end{equation}
	The inflaton $\phi$ is a real scalar field, and we consider the most general renormalizable potential\footnote{A linear term can always be removed through a field redefinition.}
	\begin{equation} \label{eq:inflaton_potential1}
		V(\phi) = b\, \phi^2 + c\, \phi^3 + d\, \phi^4.
	\end{equation}
	Additionally, we assume that DM is a Dirac fermion $\chi$ with the Lagrangian density\footnote{The results presented in this paper would be qualitatively similar for the case of a DM Majorana fermion. See Ref.~\cite{Izumine:2024hyp} for a recent study of scalar DM production during reheating.}
	\begin{equation} \label{eq:DMinflaton}
		\mathcal{L}_{\text{DM}} = i \bar{\chi}\gamma^{\mu} \partial_{\mu}\chi - \mdm\, \overline{\chi}\, \chi   -y_\chi\, \phi\, \overline{\chi}\, \chi\,,
	\end{equation}
	where $y_\chi$ denotes the Yukawa coupling to the inflaton and $\mdm$ denotes the DM mass. The reheating of the Universe is controlled by interactions of $\phi$ with the SM Higgs doublet $H$,
	\begin{equation} \label{Vphih}
		\mathcal{L}_{H\phi}= -\mu\, \phi\, H^{\dagger} H - \frac{1}{2} \lambda_{\phi H}\, \phi^2\, H^{\dagger}H  \,.
	\end{equation}
	Gravitational interactions are also relevant in our setup. Expanding the metric in Eq.~\eqref{eq:action} around a flat Minkowski background $\eta_{\mu\nu}$ as $g_{\mu \nu} \simeq \eta_{\mu \nu} + (2/M_P)\, h_{\mu \nu}$ gives rise to couplings between the energy-momentum tensor $T_i^{\mu\nu}$ for all matter fields $i$ and the graviton $h_{\mu\nu}$ as~\cite{Choi:1994ax}
	\begin{equation}\label{eq:gra}
		\sqrt{-g}\, \mathcal{L} \supset \frac{1}{M_P}\, h_{\mu \nu}\, \sum_i T_i^{\mu \nu},
	\end{equation}
	with $i =\{\phi,\, \chi,\, \text{SM}\}$. These gravitational interactions can play a crucial role in particle production in the early Universe, as we will discuss later.
	
	%%%%%%%%%%%%%%%%%%%%%%%%%%%%%%%%%%%%%%%%%%%%%%
	\section{Inflation and (P)reheating} \label{sec:inflation_reheating}
	%%%%%%%%%%%%%%%%%%%%%%%%%%%%%%%%%%%%%%%%%%%%%%
	In the following, we briefly review the dynamics of polynomial inflation and the cosmic (p)reheating taking place after inflation.
	
	%%%%%%%%%%%%%%%%%%%%%%%%%%%%%%%%%%%%%%%%%%%%%%
	\subsection{Polynomial Inflation}
\begin{figure}[t!]
		\def\sepf{0.61}
		\centering
		\includegraphics[scale=\sepf]{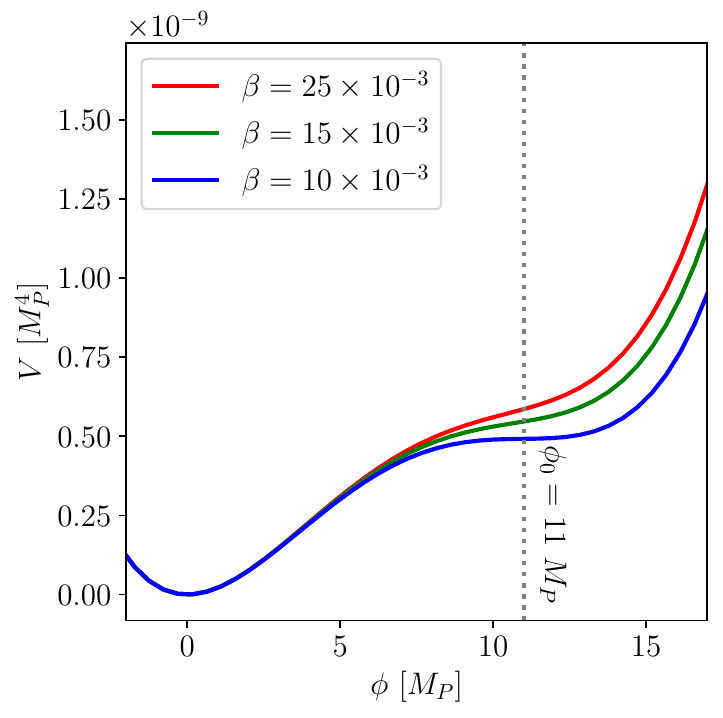}
		\caption{Inflaton potential as function of $\phi$ and $\beta$, where we have fixed $\phi_0 = 11~M_P$ and $d \simeq 10^{-13}$.}
		\label{fig:potential}
	\end{figure}

	%%%%%%%%%%%%%%%%%%%%%%%%%%%%%%%%%%%%%%%%%%%%%%
	In polynomial inflation, the potential takes the form given in Eq.~\eqref{eq:inflaton_potential1}. In the case where $b = \frac{9}{32} \frac{c^2}{d}$, the potential has an inflection point at $\phi = \phi_0 \equiv -\frac{3}{8} \frac{c}{d}$, and can be reparameterized as~\cite{Drees:2021wgd, Drees:2022aea}
	\begin{equation}  \label{inflaton_potential2}
		V(\phi) = d \left[2 \phi_0^2\, \phi^2 - \frac83 (1 - \beta)\, \phi_0\, \phi^3 + \phi^4\right].
	\end{equation}
	The parameter $\beta$ controls the flatness of the potential near $\phi_0$. For $\beta = 0$, one has an exact inflection point at $\phi_0$. A false vacuum is generated at $\phi > \phi_0$ if $\beta < 0$, in which case the inflaton may not roll down to the global minimum of the potential. The latter possibility, which is physically disfavored as it would not lead to a hot Big Bang, is avoided in this paper by only considering $0<\beta \ll 1$. To illustrate the dependence of the inflaton potential on the model parameters, we present an example in Fig.~\ref{fig:potential}. Here, we fix $\phi_0 = 11\, M_P$ (vertical dotted line) and $d \simeq 10^{-13}$. The red, green and blue curves correspond to $\beta = 25 \times 10^{-3}$, $\beta = 20 \times 10^{-3}$, and $\beta = 10 \times 10^{-3}$, respectively. As noted above, a larger $\beta$ results in a steeper potential, while a smaller $\beta$ leads to a flatter potential.
	%%%%%%%%%%%%%%%%%%%%%%%%%%%%%%%%%%%%%%%%%%%%%%
	%%%%%%%%%%%%%%%%%%%%%%%%%%%%%%%%%%%%%%%%%%%%%%
	
	We apply the SR approximation and define the following SR parameters~\cite{Lyth:2009zz}
	\begin{equation}
		\epsilon_V \equiv \frac{M_P^2}{2} \left(\frac{V^{\prime}}{V}\right)^2, \quad
		\eta_V \equiv M_P^2\, \frac{V^{\prime \prime}}{V}  \,, \quad
		\xi_V^2 \equiv M_P^4\, \frac{V^{\prime}\, V^{\prime \prime \prime}}{V^2} \,,
	\end{equation}
	which must be smaller than one during the accelerated expansion of the Universe. Inflation ends when the field reaches the value $\phi = \phiend$ defined as $\epsilon_V (\phiend) = 1$.\footnote{In large field inflation, the potential is dominated by the quadratic term at the end of inflation, which is independent of $\beta$.}  The total number of $e$-folds $N_\star$ between the time when the CMB pivot scale $k_{\star} = 0.05$~Mpc$^{-1}$ first crossed the horizon until the end of inflation is given by~\cite{Drees:2022aea}
	\begin{align}
		N_\star &= \int^{\phi_\star}_{\phiend} \frac{1}{\sqrt{2\, \epsilon_V(\phi)}}\, \frac{d\phi}{M_P} \nonumber\\
		&\simeq \frac{1}{24} \Bigg[ 3 \left(\frac{\phi}{M_P}\right)^2 - 4\, \frac{\phi\, \phi_0}{M_P^2} + 15 \left(\frac{\phi_0}{M_P}\right)^2 \nonumber\\
		& \qquad\quad - \left(\frac{\phi_0}{M_P}\right)^2 \sqrt{\frac{2}{\beta}} \arctan\left( \frac{\phi_0 -\phi }{\sqrt{2\beta}\, \phi_0} \right) - \left(\frac{\phi_0}{M_P}\right)^2 \ln\left(\frac{\phi_0 - \phi}{M_P}\right)^2 \Bigg] \Bigg|^{\phi_\star}_{\phiend}.
	\end{align}
	In this work, we consider $50 \lesssim N_\star \lesssim 65$~\cite{Liddle:2003as, Drees:2022aea}. The power spectrum of the curvature perturbation $\mathcal{P}_{\zeta}$, the spectral index $n_s$, its running $\alpha$,\footnote{The running, defined as $\alpha \equiv \frac{dn_s}{d \ln k}$, describes the scale dependence of the spectral index.} and the tensor-to-scalar ratio $r$ are respectively given by~\cite{Drees:2021wgd, Bernal:2021qrl, Drees:2022aea}
	\begin{align}
		\mathcal{P}_{\zeta} &= \frac{1}{24\pi^2\, \epsilon_V}\, \frac{V}{M_P^4}\,,\\
		n_s &= 1- 6\epsilon_V + 2\eta_V\,,\\
		\alpha &= 16\, \epsilon_V\, \eta_V - 24\, \epsilon_V^2 -2\, \xi_V^2 \,,\\
		r &= 16\, \epsilon_V,
	\end{align}
	in the SR formalism. The observables are constrained by the Planck 2018 measurements including baryonic acoustic oscillations at the pivot scale $k_{\star} = 0.05$~Mpc$^{-1}$ to~\cite{Planck:2018vyg}
	\begin{equation}  \label{planck2018}
		\mathcal{P}_{\zeta} = (2.1 \pm 0.1) \times 10^{-9}\,,\quad
		n_s =  0.9659  \pm 0.0040\,,\quad
		\alpha = -0.0041 \pm 0.0067\,.
	\end{equation} 
	The most recent constraint on $r$ is obtained from BICEP/Keck 2018 together with Planck data~\cite{BICEP:2021xfz}:
	\begin{equation} \label{BK2018}
		r< 0.035,\qquad \text{at 95\%~C.L.}
	\end{equation} 
	
	Note that the model contains four parameters that play a role in inflation, namely $d$, $\beta$, $\phi_\star$, and $\phi_0$. The value of $\phi_\star$ is fixed by setting the number of $e$-folds, $N_\star$, while the three remaining degrees of freedom must satisfy the constraints in Eqs.~\eqref{planck2018} and~\eqref{BK2018}. The parameter space allowed for $\phi_0$ is found to be~\cite{Drees:2021wgd, Drees:2022aea}
	\begin{equation}\label{eq:phi0_range}
		3\times 10^{-5}~M_P \lesssim \phi_0 \lesssim 21.5~M_P\,.
	\end{equation}
	For smaller $\phi_0$, it is required that $\phi_\star$ be close to $\phi_0$ to fit the CMB data. In this case, even a small loop correction to the potential from the inflaton couplings could affect the flatness required by CMB observations~\cite{Drees:2021wgd}. 
	On the other hand, for very large values of $\phi_0$, $\phi_\star \ll \phi_0$, the first term in Eq.~\eqref{inflaton_potential2} dominates. This implies that when $\phi_0$  is larger than a critical value, the potential becomes near quadratic, $V(\phi) \simeq 2\, d\, \phi_0^2\, \phi^2$. The latter is in tension with the upper bound on $r$ from BICEP/Keck 2018 data~\cite{BICEP:2021xfz}, which allows us to derive an upper bound on $\phi_0$.
	
	Having presented the allowed range of $\phi_0$ in Eq.~\eqref{eq:phi0_range}, we can now discuss the parameter space of the inflaton mass $m_\phi$, which is the second derivative of the potential in Eq.~\eqref{inflaton_potential2},
	\begin{equation} \label{mass}
		m_\phi = 2\sqrt{d}\, \phi_0 \simeq 5.14 \times 10^{-8} \left( \frac{\phi_0^2}{M_P} \right),
	\end{equation}
	where the last equality was obtained by using $d \simeq 6.61 \times 10^{-16}~(\phi_0/M_P)^2$.\footnote{The parameter $d$ controls the overall amplitude of the power spectrum $\mathcal{P}_{\zeta}$ and was obtained in Ref.~\cite{Drees:2021wgd} by fixing $\mathcal{P}_{\zeta}$ to the central value in Eq.~\eqref{planck2018}.} This analytical expression remains a good approximation for $\phi_0 \lesssim 5\, M_P$. For $\phi_0 \gtrsim 5\, M_P$, the range of $d$ is found to be $10^{-13} \lesssim d \lesssim 10^{-14}$~\cite{Drees:2022aea}. From the constraints on $d$ and $\phi_0$ presented here and in Eq.~\eqref{eq:phi0_range}, respectively, we find that the inflaton mass can acquire the following range of possible values
	\begin{equation}
		10^2~\text{GeV} \lesssim m_\phi \lesssim 10^{13}~\text{GeV}\,.    
	\end{equation}
	Future CMB experiments could further constrain the upper limit on the inflaton mass. The prediction for the running $\alpha$ of the spectral index is negative and lies in the range~\cite{Drees:2021wgd, Drees:2022aea}
	\begin{equation}
		\alpha  \in \left[-\mathcal{O}(10^{-2}),\, -\mathcal{O}(10^{-3})\right].
	\end{equation}
	
	%%%%%%%%%%%%%%%%%%%%%%%%%%%%%%%%%%%%%%%%%%%%%%
	\begin{figure}[t!]
		\def\sepf{0.61}
		\centering
		\includegraphics[scale=\sepf]{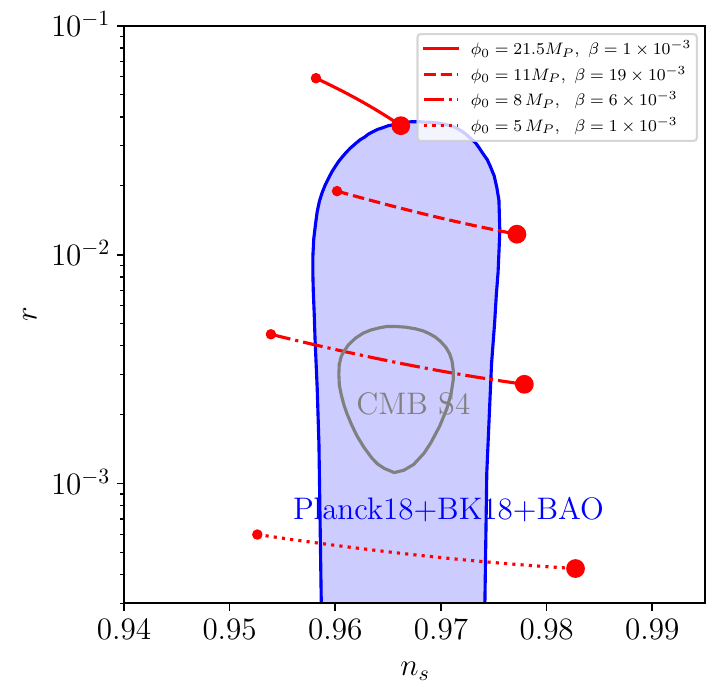}
		\includegraphics[scale=\sepf]{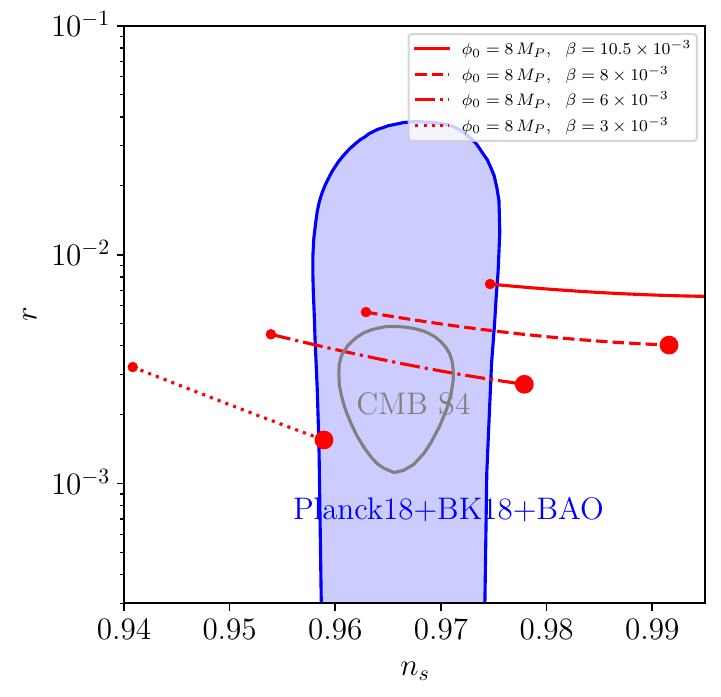}
		\caption{Left panel: Predictions in the $[n_s,\, r]$ plane for four benchmark values of $\phi_0$ and \textcolor{blue}{$\beta$}. Right panel: Impact of $\beta$ on inflationary predictions for fixed $\phi_0$.  The small (large) red dot corresponds to $N_\star =50$ ($N_\star =65$). The blue shaded area is allowed by the combination of BICEP/Keck and Planck data~\cite{BICEP:2021xfz}, while the gray line corresponds to a projected sensitivity of CMB-S4~\cite{Abazajian:2019eic}.}
		\label{fig:ns_r}
	\end{figure} 
	%%%%%%%%%%%%%%%%%%%%%%%%%%%%%%%%%%%%%%%%%%%%%%
	In the left panel of Fig.~\ref{fig:ns_r}, we show predictions in the $[n_s,\, r]$ plane for four benchmark scenarios for $50 \lesssim N_\star \lesssim 65$ indicated by a small and large red dot, respectively: $\phi_0 = 21.5~M_P$ and $\beta =1\times 10^{-3}$ (solid red), $\phi_0 = 11~M_P$ and $\beta =19\times 10^{-3}$ (red dashed), $\phi_0 = 8~M_P$ and $\beta = 6\times 10^{-3}$ (dash-dotted red) and $\phi_0 =5~M_P$ with $\beta =10^{-3}$ (dotted red). Larger $\phi_0$ corresponds to a higher inflationary scale with higher predictions for $r$. For fixed $\phi_0$, a larger $\beta$ implies a steeper potential, giving rise to a larger prediction for $r$, as shown in the right panel of Fig.~\ref{fig:ns_r}. The normalization $d$ of the potential has no impact on these results. The blue shaded area indicates the parameter space allowed by the combination of BICEP/Keck 2018 and Planck data~\cite{BICEP:2021xfz}. We note that future CMB experiments, such as CMB-S4~\cite{Abazajian:2019eic} could offer more stringent constraints on $\phi_0$, as shown in gray. As demonstrated above, polynomial inflation, depending on the parameters $\phi_0$, $d$, $\beta$, and $\phi_\star$, can well accommodate recent CMB data. In the large-field scenario, the potential is dominated by the quadratic term when inflation
	ends, which is independent of $\beta$. For the focus of this paper, namely the DM production during reheating, the relevant parameters are the inflaton mass and the inflationary scale, which both depend on $\phi_0$.
	
	%%%%%%%%%%%%%%%%%%%%%%%%%%%%%%%%%%%
	\subsection{(P)reheating} \label{sec:(p)reheating}
	%%%%%%%%%%%%%%%%%%%%%%%%%%%%%%%%%%%
	After the end of cosmic inflation, the inflaton oscillates around the minimum of its potential and transfers energy into a plasma of lighter degrees of freedom. The inflaton oscillates around $\phi= 0$ during reheating, where the quadratic term in Eq.~\eqref{inflaton_potential2} dominates, leading to $V(\phi)\simeq \frac{m_\phi^2}{2} \phi^2$. In our setup, the dominant channel for energy transfer is the decay of inflatons into pairs of Higgs boson $H$. The corresponding decay rate reads
	\begin{align} \label{eq:GammaPhi}
		\Gamma_\phi = \frac{1}{8\pi}\, \frac{\mu^2}{m_\phi} \left[1 - \left(\frac{2\, m_H}{m_\phi}\right)^2\right]^{1/2} \simeq\frac{1}{8\pi}\, \frac{\mu^2}{m_\phi}\,.
	\end{align}
	A few comments are in order before proceeding. We note that the inflaton decay rate generally receives temperature-dependent corrections due to final-state statistics, modifying it to $\Gamma_\phi(T) = \Gamma_{\rm vac} \frac{e^{m_\phi / 2T} \pm 1}{e^{m_\phi / 2T} \mp 1}$, where the upper (lower) sign corresponds to bosons (fermions) in the final state~\cite{Adshead:2019uwj}. We have verified that this temperature correction remains moderate, and thus the use of the vacuum decay rate provides a good approximation within the current setup. In addition, the coupling between the inflaton and $H$ introduces an effective mass term for $H$ proportional to $\phi$, which could alter the decay kinematics. To incorporate this effect, one must average over the inflaton oscillations, which leads to effective couplings $\mu_{\text{eff}}$ for bosonic decays. For a quadratic inflaton potential, this effect has been shown to remain relatively moderate, with $\mu_{\text{eff}} \approx \mu$~\cite{Ichikawa:2008ne, Garcia:2020wiy}. Finally, in the last step of Eq.~\eqref{eq:GammaPhi}, we neglect the effective mass of $H$ and its thermal mass correction. This remains a self-consistent and valid approximation as long as the maximum temperature obtained using Eq.~\eqref{eq:GammaPhi} satisfies $\Tmax \lesssim m_\phi$, which is the regime we consider in this work.\footnote{The thermal mass of the Higgs boson is given by $m_H^2=\left(\frac{y_t^2}{4}+\frac{3g_2^2}{16}+\frac{g_Y^2}{16}+\frac{\lambda}{2}\right)T^2$, where $y_t$, $g_2$, $g_Y$, and $\lambda$ denotes the top-Yukawa coupling, the $SU(2)$ gauge coupling, the hypercharge gauge coupling, and the quartic Higgs self-coupling, respectively~\cite{Giudice:2003jh}.} 
	
	The evolution of the inflaton energy density, $\rho_\phi$, and the radiation energy density, $\rho_R$, are governed by the following set of Boltzmann equations
	\begin{align}
		&\frac{d\rp}{dt} + 3\, \mathcal{H}\,\rp = - \Gamma_\phi\, \rp\,, \label{eq:rhophi}\\
		&\frac{d\rR}{dt} + 4\, \mathcal{H}\, \rR = + \Gamma_\phi\, \rp \label{eq:rhoR}\,,
	\end{align}
	where the Hubble expansion rate $\mathcal{H}$ is determined from the Friedmann equation
	\begin{equation}\label{eq:Hubble1}
		\mathcal{H}^2 = \frac{\rR + \rp}{3 M_P^2}\,.
	\end{equation}
	An approximate analytical solution for the SM energy density during reheating is 
	\begin{equation}\label{eq:rho_phi}
		\rp(a) \simeq \rho_\phi(\aend) \left(\frac{\aend}{a} \right)^3,
	\end{equation}
	which is valid as long as $\mathcal{H} \gg \Gamma_\phi$. Here, $\aend$ corresponds to the scale factor at the end of inflation and the beginning of reheating, $\rR(\aend) = 0$ and $\rp(\aend) = 3\, M_P^2\, \mathcal{H}^2(\aend)$. The inflationary Hubble scale $\mathcal{H}(\aend)$ is given by the Friedmann equation $\mathcal{H}^2(\aend) \simeq V(\aend)/(3M_P^2)$. 
	Using Eq.~\eqref{eq:rho_phi}, the solution for $\rR$ during reheating %from Eq.~\eqref{eq:rhoR}
	takes the form
	\begin{equation}\label{eq:rho_R_sol}
		\rR(a)=\frac{6}{5} M_P^2 \, \Gamma_\phi \, \mathcal{H}(\aend)\left(\frac{\aend}{a}\right)^\frac32 \left[1 - \left(\frac{\aend}{a}\right)^\frac52\right].
	\end{equation}
	While we have verified the validity of the above analytical approximations, we obtain our numerical results by solving the full set of Boltzmann equations without relying on any approximations.
	
	The SM radiation energy density is defined as a function of the temperature $T$ of the SM bath as\footnote{As the produced particles interact with the plasma, they eventually thermalize~\cite{Allahverdi:2010xz, Amin:2014eta}. In this work, we assume that the thermalization process occurs instantaneously, which is a good approximation if the reheating process is not dominated by Planck-suppressed operators~\cite{Harigaya:2013vwa}. To justify this, we have checked that the thermalization time scale in our framework is much shorter compared to the time scale of reheating for the benchmark model parameters under consideration. If the instantaneous-thermalization assumption is violated, then the decay products could follow initial distributions with lower occupation numbers and harder momenta~\cite{Harigaya:2013vwa, Ellis:2015jpg, Garcia:2018wtq, Chowdhury:2023jft}. }
	\begin{equation}
		\rR(T) = \frac{\pi^2}{30}\, \gs(T)\, T^4,   
	\end{equation}
	where $\gs(T)$ corresponds to the number of relativistic degrees of freedom contributing to the SM energy density. The end of reheating is determined by the onset of SM radiation domination, at a temperature $T = \Trh$. We define the reheating temperature by the equality $\mathcal{H}(\Trh) \equiv \frac23 \Gamma_\phi$, assuming that the inflaton decays away at a time scale $t= 1/\Gamma_\phi$,
	\begin{equation} \label{Trh}
		\Trh^2 = \frac {2}{\pi} \left(\frac{10}{\gs(\Trh)}\right)^\frac12 M_P\, \Gamma_\phi\,.
	\end{equation}
	
	We require $\Trh > T_\text{BBN} \simeq 4$~MeV to leave the standard BBN predictions unaffected by our setup~\cite{Sarkar:1995dd, Kawasaki:2000en, Hannestad:2004px, DeBernardis:2008zz, deSalas:2015glj}. Furthermore, the SM energy density in Eq.~\eqref{eq:rho_R_sol} has a maximum at $a =\amax \equiv (8/3)^{2/5}\, \aend$,  and the corresponding maximum temperature $\Tmax$~\cite{Giudice:2000ex} is given by
	\begin{equation}\label{eq:Tmax}
		\Tmax^4=\frac{60}{\pi^2\, \gs(\Tmax)} \left(\frac{3}{8}\right)^\frac85 M_P^2\, \Gamma_\phi\, \mathcal{H}(\aend)\,,
	\end{equation}
	which is an important parameter in the UV freeze-in process to be discussed in the next section. Finally, considering that during reheating the temperature scales as $T(a) \propto a^{-3/8}$ (cf. Eq.~\eqref{eq:rho_R_sol}), it follows that the Hubble expansion rate can be expressed as
	\begin{equation} \label{eq:H}
		\mathcal{H}(T) = \frac{\pi}{3} \, \frac{T^2}{M_P} \times
		\begin{dcases}
			\sqrt{\frac{\gs(\Trh)}{10}} \left(\frac{T}{\Trh}\right)^2 &\text{ for } \Tmax > T\geq \Trh\,,\\
			\sqrt{\frac{\gs(T)}{10}} &\text{ for } \Trh \geq T\,.
		\end{dcases}
	\end{equation}
	This result will later be used to analytically solve the Boltzmann equation governing DM production.
	
	%%%%%%%%%%%%%%%%%%%%%%%%%%%%%%%%%%%%%%
	\begin{figure}[t!]
		\def\sepf{0.5}
		\centering
		\includegraphics[scale=\sepf]{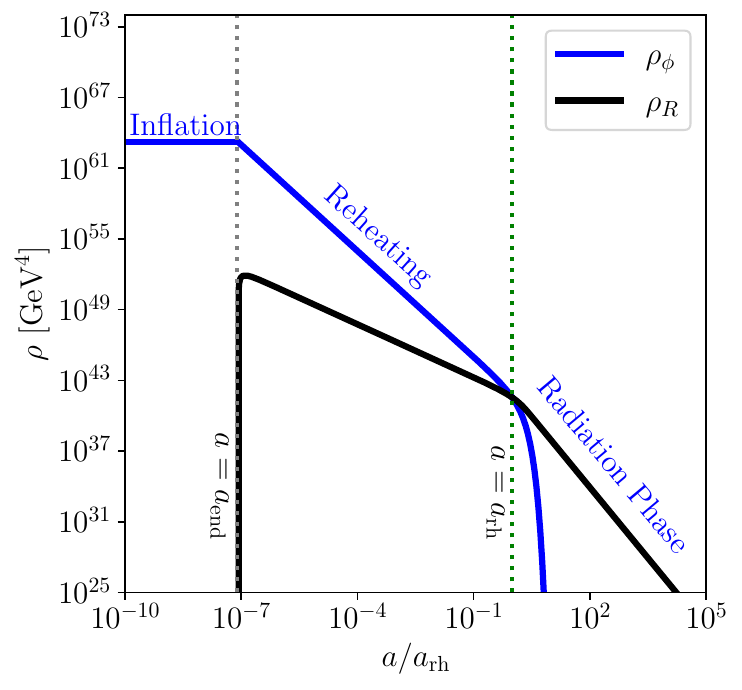}
		\includegraphics[scale=\sepf]{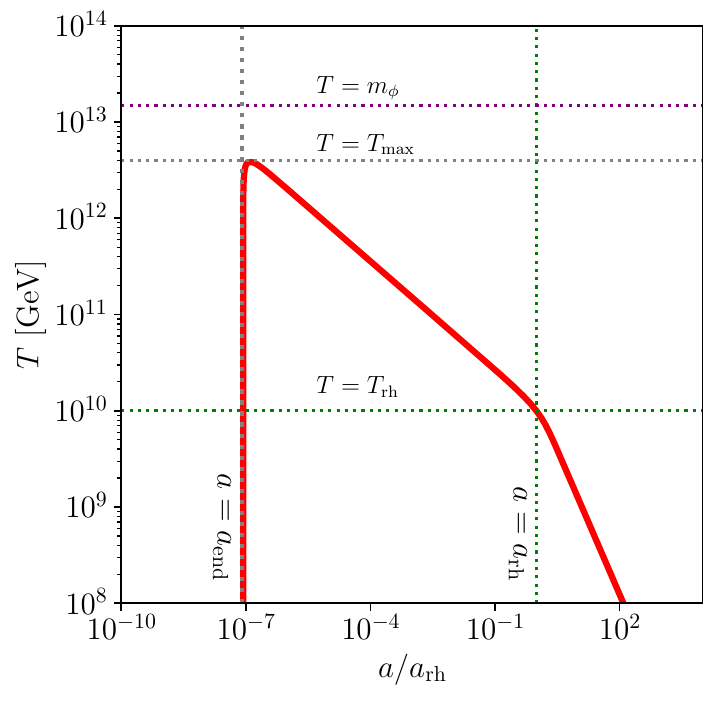}
		\caption {Evolution of energy densities (left) and SM temperature (right) for $\Trh =10^{10}$~GeV, $\phi_0 =21.5~M_P$, and $m_\phi \simeq 1.5 \times 10^{13}$~GeV.} 
		\label{fig:rho_tem}
	\end{figure} 
	%%%%%%%%%%%%%%%%%%%%%%%%%%%%%%%%%%%%%%
	To illustrate the preceding discussion, in Fig.~\ref{fig:rho_tem} we depict the evolution of the radiation and inflaton energy densities (left) and the SM temperature (right), for the largest possible value $\phi_0 =21.5~M_P$ and  $m_\phi \simeq 1.5 \times 10^{13}$~GeV (the latter is obtained from Eq.~\eqref{mass} using $d\simeq 2\times 10^{-14}$). We have numerically solved the coupled Boltzmann equations~\eqref{eq:rhophi} and~\eqref{eq:rhoR} for an illustrative scenario where $\Trh =10^{10}$~GeV. The figure shows that the maximum temperature during reheating can be much higher than $\Trh$~\cite{Giudice:2000ex}, which has important implications for the production of DM discussed in the following sections. For a fixed reheating temperature, the inflaton decay rate is obtained via $\Gamma_\phi = 3/2\, \mathcal{H}(\Trh)$ with $\mathcal{H}(\Trh) =  \frac{\pi}{3} \sqrt{\frac{\gs(\Trh)}{10}}\, \frac{\Trh^2}{M_P} $. This means that the inflaton does not thermalize during reheating, as the inflaton interaction rate $\Gamma_\phi$ remains smaller than the Hubble rate throughout reheating.
	
	In general, inflaton couplings to other particles could spoil the flatness of the potential during inflation. In the polynomial inflationary setup, the radiative stability of the inflaton potential during inflation requires a suppressed inflaton-Higgs coupling, which in turn requires~\cite{Drees:2021wgd}
	\begin{equation} \label{eq:mubound}
		\left| \left( \frac{ \mu} {\phi_0} \right)^2 \ln \left( \frac{\mu} {\phi_0} \right) - \left( \frac {\mu} {\phi_0}\right)^2 \right| < 64\pi^2 d\, \beta\,,
	\end{equation}
	and translates into an upper bound on the coupling $\mu$. The constraint on $\mu$ shown in Eq.~\eqref{eq:mubound} and the upper bound on $\phi_0=21.5M_{P}$ in Eq.~\eqref{eq:phi0_range} gives rise to an upper bound on the maximal reheating temperature $   \Trh \lesssim  10^{14}~\text{GeV} $  in a general setup ~\cite{Drees:2022aea}. In the reheating formalism employed in this work, which is valid for $\Tmax\lesssim m_\phi$ (cf. Eq.~\eqref{eq:GammaPhi}), we obtain the following constraint on the reheating temperature from the consistency of our calculation
	\begin{equation} \label{eq:Trh_bound}
		\Trh \lesssim 1.8 \times 10^{11}~\text{GeV}. 
	\end{equation}
	We emphasize that the upper bound on $\Trh$ depends on $\phi_0$ and the couplings involved for the reheating.  
	Similarly, DM would also contribute (negatively) to the inflaton potential at loop level, which could potentially spoil inflationary predictions. To keep radiative corrections under control, $y_{\chi}$ must satisfy~\cite{Drees:2021wgd}
	\begin{equation}
		\left|y_{\chi}^4  - 3\, y_{\chi}^4\, \ln\left(y_{\chi}^2\right)\right| < 64 \pi^2\, d\,\beta\,.
	\end{equation}
	Assuming  the  upper bound on $\phi_0=21.5M_{P}$ and taking into account the allowed range of $d$ and $\beta$  within the $2 \sigma$ range of the BICEP 2018 data, we obtain
	\begin{equation} \label{eq:ychibound}
		y_\chi \lesssim 10^{-4}\,.  
	\end{equation}
	We note that within our framework, non-perturbative preheating and pure gravitational reheating are subdominant to perturbative decay. The trilinear coupling, $\mu\, \phi\, |H|^2$, which induces a tachyonic squared-mass $m^2_{H} \sim \mu\, \phi$ for $H$ once $\phi$ becomes negative after crossing zero, tends to make preheating efficient. However, the quartic self-interaction term, $\lambda|H|^4$, introduces a positive squared mass term $m^2_{H} \sim \lambda \left\langle H^2 \right \rangle $, where $\left\langle H^2 \right \rangle $ denotes the variance of $H$. This backreaction effectively counteracts the tachyonic effects, quickly terminating preheating and rendering it inefficient~\cite{Dufaux:2006ee}. Finally, we comment on the possibility of gravitational reheating. It has been established that steep inflaton potentials $V\sim \phi^p$ with $p > 9$ are required at the reheating stage for successful gravitational reheating~\cite{Clery:2021bwz, Haque:2022kez, Barman:2022qgt,  Haque:2023yra}. Interestingly, this bound can be relaxed to $p > 4$ by introducing right-handed neutrinos~\cite{Haque:2023zhb} or a non-minimal coupling between gravity and two inflatons~\cite{Clery:2022wib}, or even to $p \geq 2$ by coupling gravity non-minimally to a single inflaton~\cite{Barman:2023opy}. In our case, the inflaton oscillates around a quadratic potential $V(\phi) \propto \frac12\, m_\phi^2\, \phi^2$ during reheating and is minimally coupled to gravity, making gravitational reheating inefficient. It is therefore ignored in the following.
	
	%%%%%%%%%%%%%%%%%%%%%%%%%%%%%%%%%%%%%%%%%%%%%%%%%%%%%%%%%
	\section{Dark Matter Production} \label{sec:dm}
	%%%%%%%%%%%%%%%%%%%%%%%%%%%%%%%%%%%%%%%%%%%%%%%%%%%%%%%%%
	In the following, we aim to quantify DM production within the large-field inflationary model during and after reheating. The evolution of the DM number density $n$ can be tracked using the Boltzmann equation
	\begin{equation} \label{eq:DMBE01}
		\frac{dn}{dt} + 3\, \mathcal{H} \,n  = \gamma\,, 
	\end{equation}
	where $\gamma$ denotes the DM production rate density. Taking into account that during reheating the SM entropy is not conserved, it is convenient to introduce a comoving number density $N \equiv n\, a^3$, with which the Boltzmann Eq.~\eqref{eq:DMBE01} can be rewritten as 
	\begin{equation} \label{eq:DMBE02}
		\frac{dN}{da} = \frac{a^2\, \gamma}{\mathcal{H}}\,,
	\end{equation}
	or equivalently as
	\begin{equation} \label{eq:BEduring}
		\frac{dN}{dT} =
		\begin{dcases}
			-\frac{8}{\pi} \sqrt{\frac{10}{\gs(\Trh)}}\, \frac{M_P\, \Trh^{10}}{T^{13}} a^3(\Trh)\, \gamma &\text{ for } \Tmax > T \geq \Trh\,,\\
			-\frac{3}{\pi} \sqrt{\frac{10}{\gs(T)}}\, \frac{\gss(\Trh)}{\gss(T)}\, \frac{M_P\, \Trh^3}{T^6}\, a^3(\Trh)\, \gamma &\text{ for } \Trh \geq T\,,
		\end{dcases}
	\end{equation}
	using Eq.~\eqref{eq:H}. We solve Eq.~\eqref{eq:BEduring} to obtain the number density. To facilitate a comparison with observational data, we introduce the DM yield $Y(T) \equiv n(T)/s(T)$ as a function of the SM entropy density $s(T) \equiv \frac{2\pi^2}{45}\, \gss\, T^3$, where $\gss(T)$ is the number of relativistic degrees of freedom that contribute to the SM entropy. 
	To fit the observed relic abundance, the DM yield has been fixed such that 
	\begin{equation}
		\mdm\,Y_0 = \Omega_\chi h^2 \, \frac{1}{s_0}\,\frac{\rho_c}{h^2} \simeq 4.3 \times 10^{-10}~\text{GeV}\,,
	\end{equation}
	with $\rho_c \simeq 1.05 \times 10^{-5} \, h^2$~GeV/cm$^3$ the critical energy density of the Universe, $s_0\simeq 2.9\times 10^3$~cm$^{-3}$ the present entropy density~\cite{ParticleDataGroup:2020ssz}, and $\Omega_\chi h^2 \simeq 0.12$~\cite{Planck:2018vyg}.
	
	%%%%%%%%%%%%%%%%%%%%%%%%%%%%%%%%%%%%%%%%%%%%%%
	\begin{figure}[t!]
		\centering
		\includegraphics[scale=0.15]{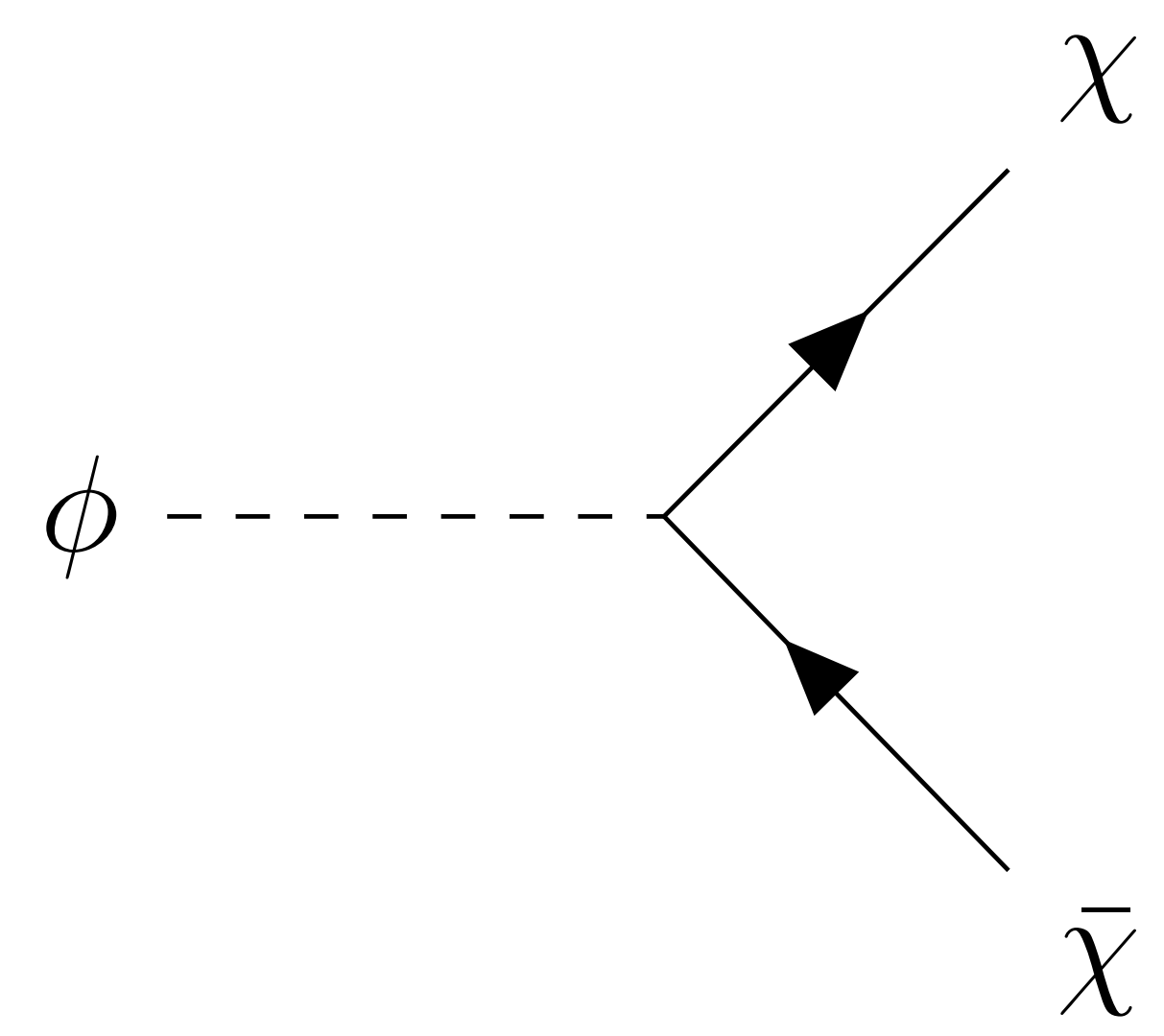}\\
		\includegraphics[scale=0.15]{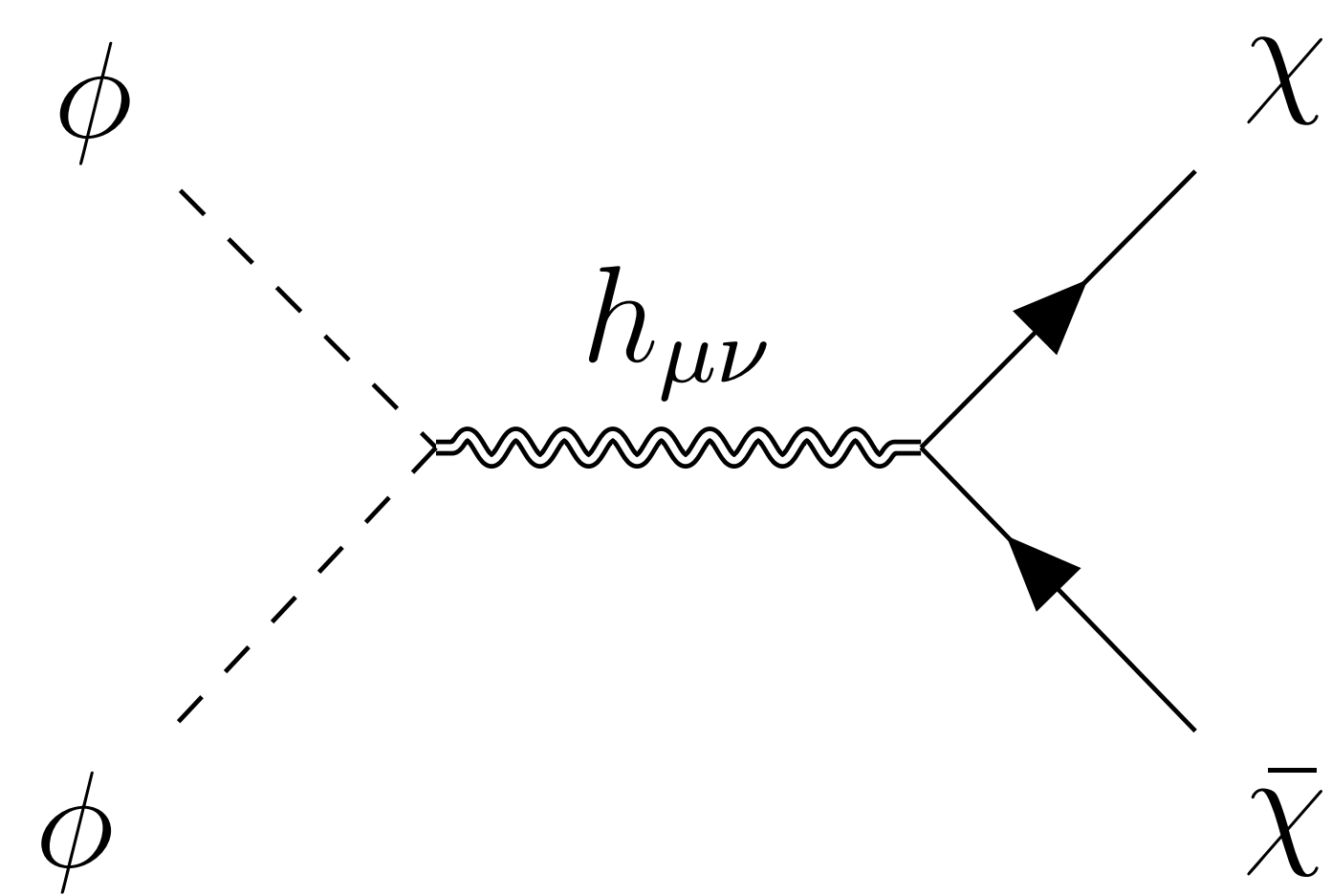}
		\includegraphics[scale=0.15]{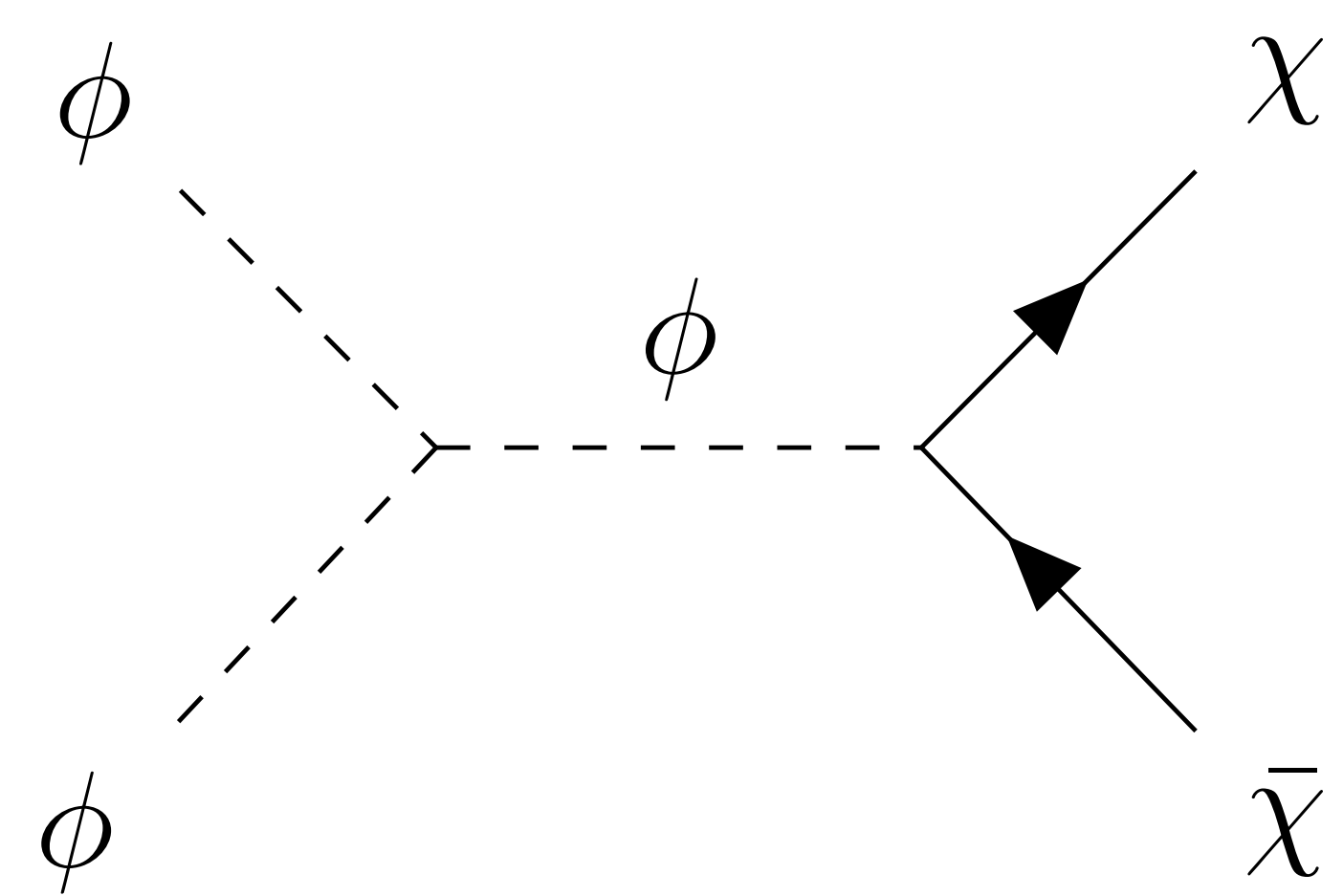}
		\includegraphics[scale=0.15]{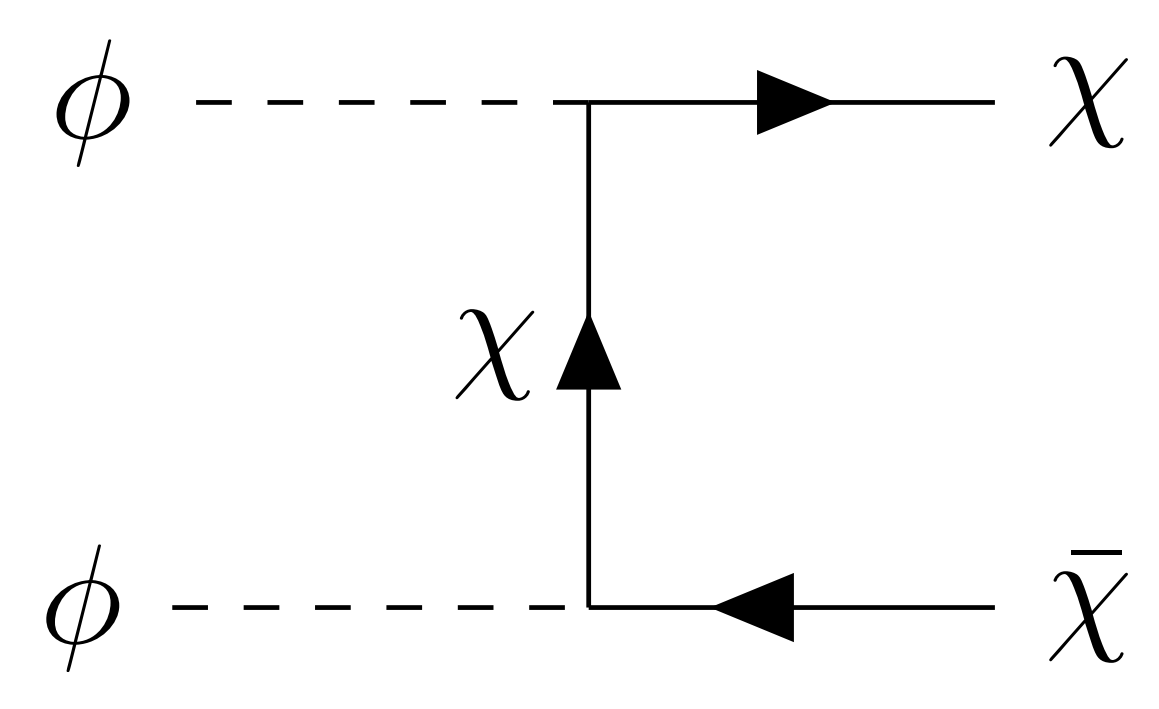}\\
		\includegraphics[scale=0.15]{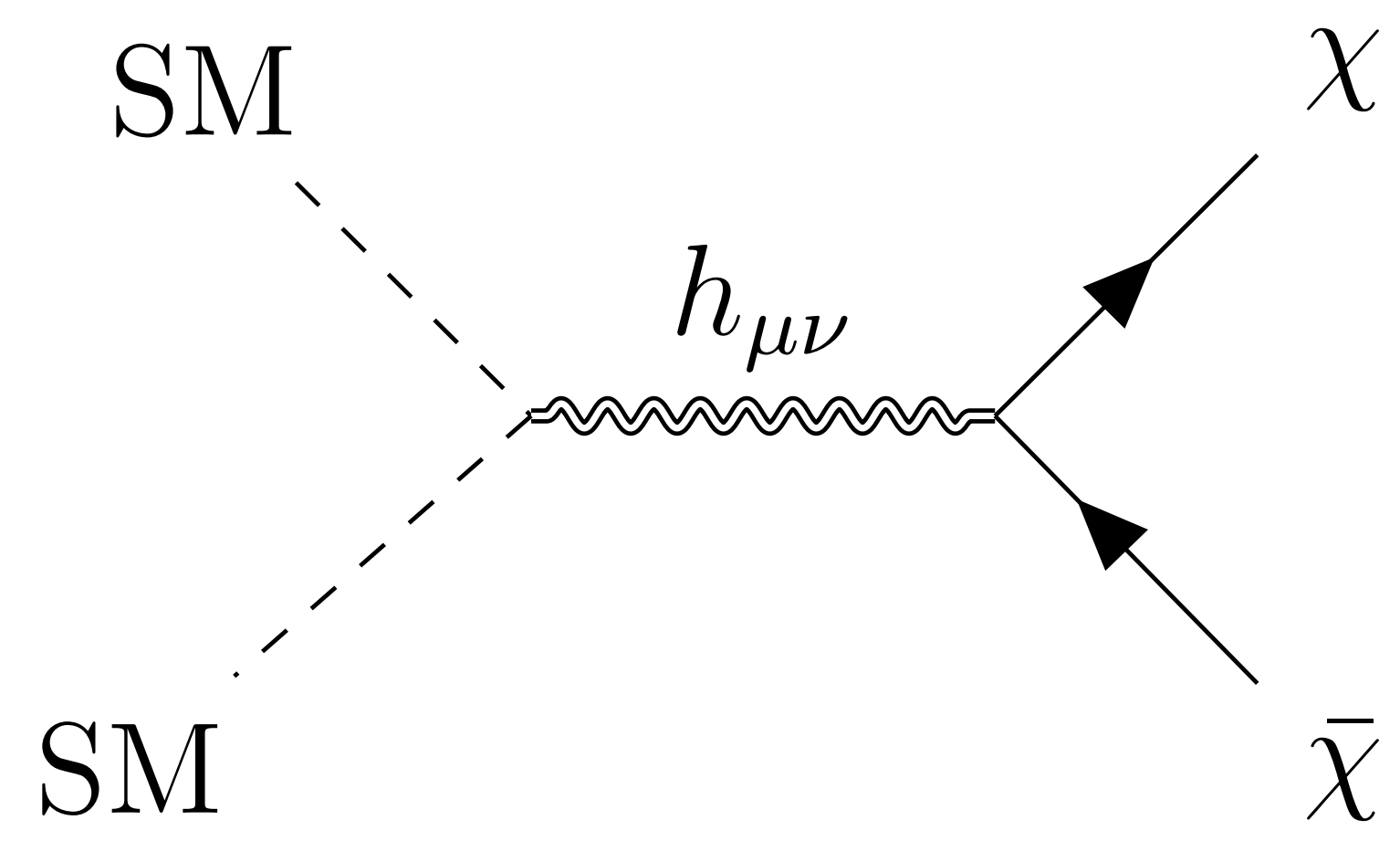}
		\includegraphics[scale=0.15]{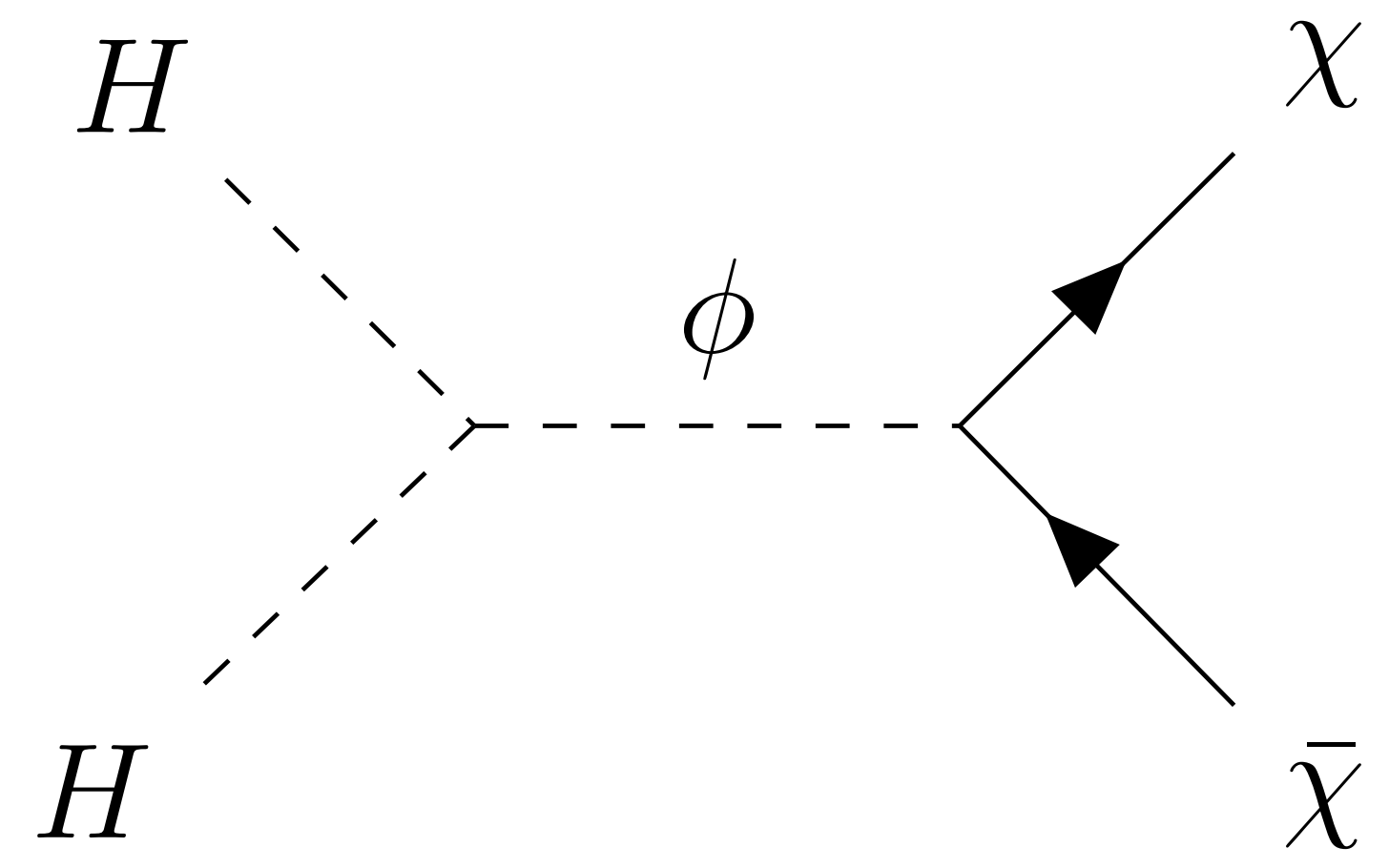}
		\caption{Feynman diagram for fermionic DM production.}
		\label{fig:dia_fermion}
	\end{figure} 
	%%%%%%%%%%%%%%%%%%%%%%%%%%%%%%%%%%%%%%%%%%%%%%
	The relevant tree-level processes that contribute to DM production are depicted in Fig.~\ref{fig:dia_fermion}, and include 
	\begin{itemize}
		\item Inflaton decay (first row),
		\item Inflaton scattering (second row),
		\item SM particle scattering (third row),
	\end{itemize}
	which we will investigate in the following three subsections, respectively.
	
	%%%%%%%%%%%%%%%%%%%%%%%%%%%%%%%%%%%%%%%%%%%%%%
	\subsection{Inflaton Decay}
	\label{subsec:InflatonDecays}
	%%%%%%%%%%%%%%%%%%%%%%%%%%%%%%%%%%%%%%%%%%%%%%
	The interaction rate density associated with the production of pairs of DM particles from inflaton decays (through the trilinear interaction in Eq.~\eqref{eq:DMinflaton}) is given by
	\begin{equation}\label{eq:decay_gamma}
		\gamma = 2\, n_{\phi}\, \Gamma_\phi\, \text{Br}\,,
	\end{equation}
	where $n_{\phi} =\rho_\phi/m_\phi$ denotes the inflaton number density, and $\Gamma_\phi$ is given in Eq.~\eqref{eq:GammaPhi}. Note that two DM particles are produced in each inflaton decay, which is accounted for by the factor 2 in Eq.~\eqref{eq:decay_gamma}. Additionally, the branching ratio, $\text{Br}$, for the decay of the inflaton into DM is given by 
	\begin{equation}\label{eq:BR}
		\text{Br} \equiv \frac{\Gamma_{\phi \to \bar{\chi} \chi}}{\Gamma_{\phi \to H^{\dagger}H} + \Gamma_{\phi \to \bar{\chi} \chi}} \simeq y_\chi^2 \left(\frac{m_\phi}{\mu}\right)^2 \left(1-\frac{4\, \mdm^2}{m_{\phi}^2}\right)^{3/2}.
	\end{equation}
	With the interaction rate in Eq.~\eqref{eq:decay_gamma}, one can solve Eq.~\eqref{eq:BEduring} during reheating, i.e. for $\Tmax \geq T \geq \Trh$, and obtain
	\begin{equation}
		N(\arh) \equiv   N(\Trh) \ \simeq \frac{2\pi\, \gs(\Trh)}{15} \sqrt{\frac{10}{\gs(\Trh)}}\, \text{Br}\, \frac{M_P\, \Trh^2\, \Gamma_\phi}{m_\phi}\, \arh^3\,,
	\end{equation}
	from which we arrive at the DM yield at present
	\begin{equation} \label{Y0Decay}
		Y_0=Y_{\text{rh}} = \frac{N(\arh)}{s(\Trh)\, \arh^3} \simeq \frac32\, \frac{\gs(\Trh)}{\gss(\Trh)}\, \frac{\Trh}{m_\phi}\, \text{Br} \simeq \frac{3\, y_\chi^2}{8\pi^2}\, \frac{\sqrt{10\, \gs(\Trh)}}{\gss(\Trh)}\, \frac{M_P}{\Trh}\,.
	\end{equation}
	Note that DM production from inflaton decay is exponentially suppressed after reheating and can be neglected.
	
	%%%%%%%%%%%%%%%%%%%%%%%%%%%%%%%%%%%%%%%%%%%%%%%
	\begin{figure}[t!]
		\def\sepf{0.5}
		\centering
		\includegraphics[scale=\sepf]{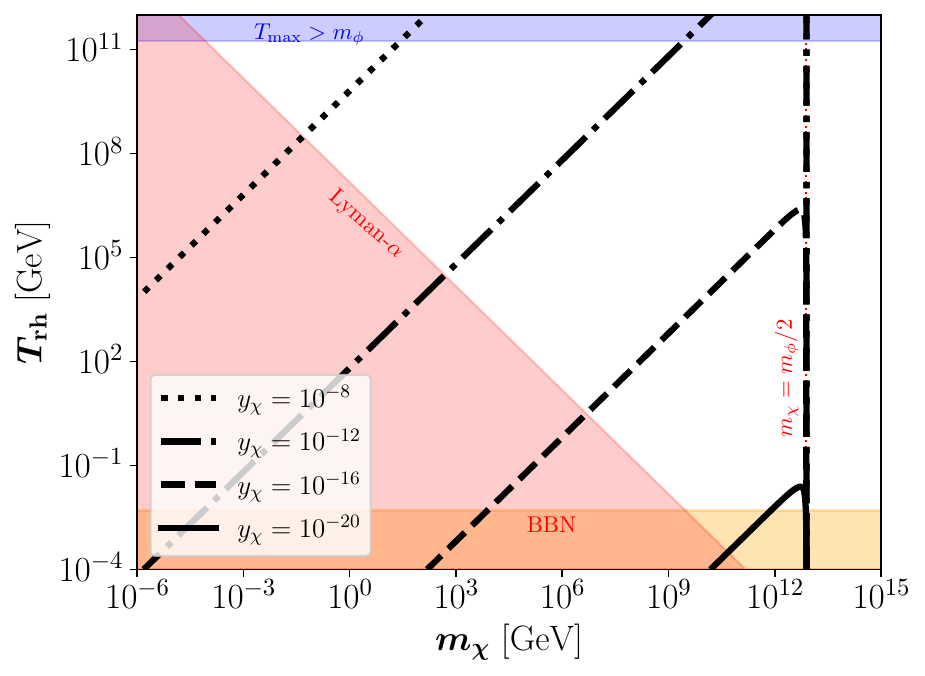}
		\caption{Parameter space that reproduces the observed DM abundance for DM produced through 2-body decays of the inflaton for different values of the Yukawa coupling $y_\chi$ with $m_\phi = 1.5\times 10^{13}~\text{GeV}$ and $\phi_0 = 21.5~M_P$  (thick black lines). The colored bands are in tension with BBN (orange), Lyman-$\alpha$ (red) or generate radiative corrections from $H$ particle (blue) or DM loops (green) to the potential that could spoil the flatness required for successful inflation. The vertical red dotted line corresponds to $\mdm = m_\phi/2$.}
		\label{fig:decay}
	\end{figure} 
	%%%%%%%%%%%%%%%%%%%%%%%%%%%%%%%%%%%%%%%%%%%%%%%
	In Fig.~\ref{fig:decay}, we show with thick black lines the parameter space $[\mdm,\, \Trh]$ required to reproduce the observed DM relic density through inflaton decays, for several benchmark values for the Yukawa coupling $y_\chi$. As an example, we have again fixed $\phi_0 = 21.5~M_P$, which 
	corresponds to an inflaton mass $m_\phi \simeq 1.5 \times 10^{13}$~GeV for $d = 2\times 10^{-14}$. Energy conservation dictates that only DM with masses up to $m_\phi/2$ can be produced from inflaton decays, as illustrated by the vertical red line in Fig.~\ref{fig:decay}. Several constraints to the model parameters are overlaid in color. Low reheating temperatures with $\Trh \lesssim 4~\text{MeV}$~\cite{Sarkar:1995dd,  Kawasaki:2000en, Hannestad:2004px, DeBernardis:2008zz, deSalas:2015glj} are in tension with BBN, whereas high reheating temperatures are constrained by $\Tmax\lesssim m_\phi$. Finally, small masses are not allowed due to the Lyman-$\alpha$ constraints: The DM particles produced from inflaton decays had an initial energy $E_\chi = m_\phi/2$ at the time of production. Due to redshifting, the present momentum of the DM particles is $p_\chi(a_0) = E_\chi \times \arh/a_0$ with $a_0$ and $\arh$ being the scale factors at present and at the end of reheating, respectively. As a cold DM candidate, it is required that the DM velocity at present is smaller than that of warm DM constrained by Lyman-$\alpha$, namely $p_\chi(a_0)/m_\chi \lesssim 1.8 \times 10^{-8}$~\cite{Masina:2020xhk},
	which leads to~\cite{Bernal:2021qrl}
	\begin{equation}
		\frac{\mdm}{\text{keV}} \gtrsim \frac{m_\phi}{\Trh}\,.
	\end{equation}
	The Lyman-$\alpha$ constraint and the condition $y_\chi <10^{-4}$ can be combined to give a lower bound on the DM mass
	\begin{equation}
		\mdm \gtrsim 4.8~\text{keV}.
	\end{equation}
	
	We note that compared to the previous analysis in a small-field model~\cite{Bernal:2021qrl}, where $\mdm \lesssim 10^{11}$~GeV, DM can be as large as $\mdm \simeq m_\phi/2 \sim 10^{13}$~GeV in the large-field scenario. However, other production channels, e.g. gravitational production, can also be efficient in the large-field scenario and will modify the results presented in Fig.~\ref{fig:decay}, as we demonstrate below.
	
	%%%%%%%%%%%%%%%%%%%%%%%%%%%%%%%%%%%%%%%%%%%%%%%%%%%%%%%%%%%5
	\subsection{Inflaton Scattering}
	\label{subsec:InflatonScattering}
	%%%%%%%%%%%%%%%%%%%%%%%%%%%%%%%%%%%%%%%%%%%%%%%%%%%%%%%%%%%5
	In addition to the production from inflaton decays, DM particles are also generated from inflaton scatterings, as shown in the second row of Fig.~\ref{fig:dia_fermion}. The relevant processes are $s$-channel annihilations mediated by a graviton or an inflaton (due to the $\phi^3$ term in the potential) and $t$-channel annihilations mediated by DM. Compared to decays, the inflaton-mediated $s$-channel and DM-mediated $t$-channel annihilations are always suppressed because of extra couplings appearing in the cross section. We thus only focus on gravitational annihilation, which dominates the production if the direct coupling between the inflaton and the DM is sufficiently suppressed, see e.g. Refs.~\cite{Ema:2015dka, Ema:2016hlw, Ema:2018ucl, Haque:2021mab} for some related ideas.
	
\begin{figure}[t!]
		\def\sepf{0.47}
		\centering
		\includegraphics[scale=\sepf]{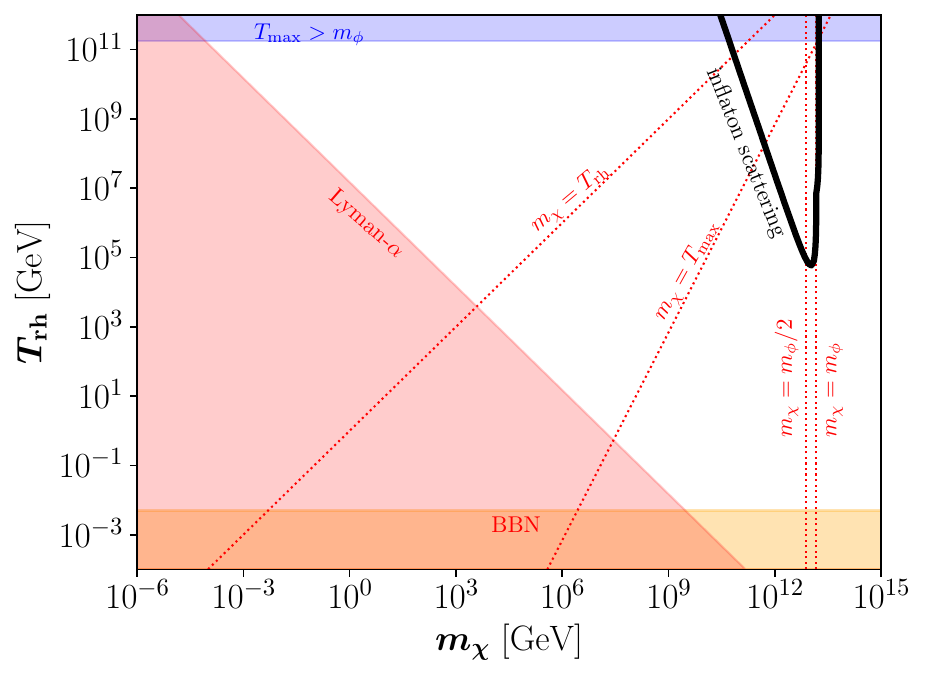}
		\caption{Parameter space that reproduces the observed DM abundance for DM produced through gravitational scatterings of inflatons or SM particles with $m_\phi = 1.5\times 10^{13}~\text{GeV}$ and $\phi_0 = 21.5M_P$  (thick black lines). The colored bands are in tension with BBN (orange), Lyman-$\alpha$ (red) or generate a radiative-unstable potential (blue). The vertical red dotted lines correspond to $\mdm = m_\phi/2$, $\mdm = m_\phi$,  $\mdm = \Trh$ and  $\mdm = \Tmax$.} 
		\label{fig:decay2}
	\end{figure}

	For the DM production from gravitational annihilation of inflatons, the interaction-rate density is given by~\cite{Mambrini:2021zpp, Bernal:2021kaj, Barman:2021ugy, Barman:2022tzk}
	\begin{equation}\label{eq:gamma_phiphichichi}
		\gamma = \frac{\rho_{\phi}^2}{m_{\phi}^2}\, \frac{ \mdm^2 }{64\pi M_P^4} \left(1-\frac{\mdm^2}{m_{\phi}^2}\right)^{3/2},
	\end{equation}
	where $\mdm$ in the second term arises from spin summation in the final states. A detailed derivation of Eq.~\eqref{eq:gamma_phiphichichi} can be found in appendix \ref{eq:appendixA}. The corresponding DM yield at present is
	\begin{equation} \label{eq:Y0_gra_phiphi}
		Y_0 \simeq \frac{\gs}{128\,\gss} \sqrt{\frac{\gs}{10}} \left(\frac{\mdm}{m_{\phi}}\right)^2 \left(\frac{\Trh}{M_P}\right)^3 \left[\left(\frac{\Tmax}{\Trh}\right)^4-1\right] \left(1-\frac{\mdm^2}{m_{\phi}^2}\right)^{3/2}.
	\end{equation}
	Note that the enhancement factor $\left[(\Tmax/\Trh)^4-1\right]$ arises from the noninstantaneous nature of reheating. 
	In large field polynomial inflation, a large ratio $\Tmax/\Trh$ naturally occurs (cf. Fig.~\ref{fig:rho_tem}), and the corresponding large enhancement of gravitational DM production from inflaton annihilation can successfully account for the observed relic abundance, as depicted by the solid black line in Fig.~\ref{fig:decay2}. For inflaton masses of the order $m_\phi \sim 10^{13}$~GeV, gravitational production is efficient even when the reheating temperature is as low as $\Trh \sim 10^5$~GeV in the regime where the mass of the DM reaches its kinematic threshold. Note that when $\mdm = m_\phi$, the production is closed from the phase space. By comparing Eq.~\eqref{eq:Y0_gra_phiphi} and Eq.~\eqref{Y0Decay}, we find that for a given reheating temperature $\Trh$, inflaton mass $m_\phi$, and DM mass $\mdm (\lesssim m_\phi/2)$, the gravitational annihilation dominates over the decay if
	\begin{equation} \label{eq:ychi_bound_lower}
		y_\chi \lesssim \frac{\pi}{4}\, \sqrt{\frac{\gs}{30}}\, \frac{\mdm}{m_\phi} \left(\frac{\Tmax}{M_P}\right)^2.
	\end{equation}
	%%%%%%%%%%%%%%%%%%%%%%%%%%%%%%%%%%%%%%%%%%%%%%%
	%%%%%%%%%%%%%%%%%%%%%%%%%%%%%%%%%%%%%%%%%%%%%%%
	
	\subsection{SM Particles Scattering}
	%%%%%%%%%%%%%%%%%%%%%%%%%%%%%%%%%%%%%%%%%%%%%%%%%%
	In the following, we consider DM production from scatterings of SM particles (last row of Fig.~\ref{fig:dia_fermion}), mediated by either the graviton or inflaton. These processes allow the production of DM with $\mdm \sim \Tmax$ and feature a UV freeze-in behavior due to the suppression by the Planck scale.
	
	%%%%%%%%%%%%%%%%%%%%%%%%%%%%%%%%%%%%%%%%%%%%%%%%%%
	\subsubsection{Graviton Mediation}
	%%%%%%%%%%%%%%%%%%%%%%%%%%%%%%%%%%%%%%%%%%%%%%%%%%
	For gravitational DM production from the annihilation of SM particles in the thermal plasma, the interaction rate is given by~\cite{Garny:2015sjg, Tang:2017hvq, Garny:2017kha, Bernal:2018qlk}
	\begin{equation}
		\gamma \simeq \alpha_f \frac{T^8}{M_P^4}\,,
	\end{equation}
	where $ \alpha_f\simeq 1.1\times 10^{-3}$ is obtained for fermionic DM by summing over the contributions from all SM degrees of freedom in the thermal plasma~\cite{Garny:2017kha}.
	The corresponding DM yield at present is given by
	\begin{equation} \label{smg2}
		Y_0 \simeq
		\begin{dcases}
			\frac{45\, \alpha_f}{2\pi^3\gss}\sqrt{\frac{10}{\gs}}\frac{\Trh^3}{M_P^3}\left\{\left[1-\left(\frac{\mdm}{\Trh}\right)^3\right]+2\left[1-\left(\frac{\Trh}{\Tmax}\right)^4\right]\right\} \simeq \frac{135\, \alpha_f}{2\pi^3\gss}\sqrt{\frac{10}{\gs}}\frac{\Trh^3}{M_P^3}\\
			\hspace{8.4cm} \text{for } \mdm \lesssim \Trh\,,\\
			\frac{45\, \alpha_f}{\pi^3 \gss} \sqrt{\frac{10}{\gs}}\, \frac{\Trh^7}{M_P^3\, \mdm^4} \left[ 1- \left(\frac{\mdm}{\Tmax}\right)^4\right] \qquad \qquad \qquad \text{for } \Trh \lesssim \mdm \lesssim \Tmax\,.
		\end{dcases}
	\end{equation}
	DM lighter than $\Trh$ is produced both during and after reheating, while DM heavier than $\Trh$ (and lighter than $\Tmax$) is produced only during reheating. We find that the gravitational scattering of SM particles is subdominant compared to the channels studied in the two preceding subsections.  
	
	%%%%%%%%%%%%%%%%%%%%%%%%%%%%%%%%%%%%%%%%%%%%%
	\subsubsection{Inflaton Mediation}
	%%%%%%%%%%%%%%%%%%%%%%%%%%%%%%%%%%%%%%%%%%%%%
	Finally, we focus on the inflaton-mediated UV freeze-in production of DM from scatterings of SM particles, as depicted in the second diagram of the last row of Fig.~\ref{fig:dia_fermion}. The annihilation cross section $\sigma$ for this process is given by
	\begin{equation} \label{crossection}
		\sigma(\mathfrak{s}) \simeq  \frac{1}{8\pi} \frac{y^2_{\chi}\, \mu^2}{(\mathfrak{s} -m_\phi^2)^2 + m_\phi^2\, \Gamma_\phi^2} \left(1-\frac{4\mdm^2}{\mathfrak{s}}\right)^{3/2}, 
	\end{equation}
	where $\sqrt{\mathfrak{s}}$ corresponds to the center-of-mass energy.
	\footnote{We take the Higgs to be massless before the onset of the electroweak phase transition. The thermal mass of the Higgs is dominated by $\sim \frac12\, y_t\, T$ with $y_t$ being the quark-top Yukawa coupling~\cite{Quiros:1999jp, Giudice:2003jh}, which is negligible compared to the momentum of the thermal Higgs.} 
	Since the SM temperature can be higher than the inflaton mass, the DM can be produced resonantly during reheating~\cite{Barman:2024mqo}. The corresponding interaction rate density is 
	\begin{equation} \label{eq:gamma_chichiHH}
		\gamma(T) \simeq \frac{T}{8\pi^4} \int_{4\, \mdm^2}^\infty d\mathfrak{s}\, \mathfrak{s}^{3/2}\, \sigma(\mathfrak{s})\, K_1\left(\frac{\sqrt{\mathfrak{s}}}{T}\right),
	\end{equation}
	and is well approximated by~\cite{Bernal:2020fvw, Bernal:2020yqg}
	\begin{equation} \label{gamma}
		\gamma(T) \simeq \frac{y^2_{\chi}\, \mu^2}{64 \pi^4} \times
		\begin{cases}
			\frac{32}{\pi} \frac{T^6}{m_\phi^4} &\quad \text{ for }  T \ll m_\phi\,,\\
			\frac{m_\phi^2\, T}{\Gamma_\phi}\, K_1\left(\frac{m_\phi}{T}\right) \left[1 - \left(\frac{2 \mdm}{m_\phi}\right)^2\right]^{3/2}
			&\quad \text{ for } T \gtrsim m_\phi\,,
		\end{cases}
	\end{equation}
	where $K_1$ denotes the modified Bessel function of the first kind. The analytical approximations shown in Eq.~\eqref{gamma} have been validated against the full numerical result. Finally, as a consistency check, we note that DM does not thermalize, as $\gamma(T) \ll \mathcal{H}(T)\, n_\text{eq}(T)$ with $n_\text{eq}(T)\simeq \frac{4}{\pi^2}\, T^3$ for $T \gg \mdm$.
	
	%The inflaton mass is typically larger than the reheating temperature due to the radiative stability condition~\cite{Drees:2021wgd, Bernal:2021qrl}. 
	In the case where $m_\phi \gg \Trh$, which is a good approximation due to the requirement $\Tmax\lesssim m_\phi$, we obtain the following analytic approximation for the DM yield produced from inflaton-mediated Higgs scattering using the first line of Eq.~\eqref{gamma},\footnote{Equation~\eqref{eq:Y0_1} refines the result reported in Ref.~\cite{Bernal:2021qrl} by a factor $195/135$ accounting for the DM production during reheating, which was not considered previously.}
	\begin{equation} \label{eq:Y0_1}
		Y_0 \simeq \frac{195\, y_\chi^2}{4\pi^8\, \gss}\, \sqrt{\frac{10}{\gs}}\, \frac{\mu^2\, \Trh\, M_P}{m_\phi^4}\, \simeq \frac{195\, y_\chi^2 }{\gss \, \pi^6} \left(\frac{\Trh}{m_\phi}\right)^3\,.
	\end{equation}
	This result is parametrically suppressed by a factor 
	\begin{equation} \label{smallfieldsupp}
		\frac{\Trh}{M_P} \left(\frac{\Trh}{m_\phi}\right)^3 \ll 1
	\end{equation}
	compared to the contribution from inflaton decays (cf. Eq.~\eqref{Y0Decay}), and hence always negligible. 
	
	%%%%%%%%%%%%%%%%%%%%%%%%%%%%%%%%%%%%%%%%%
	\subsection{Combined analysis}
\begin{figure}[t!]
		\def\sepf{0.31}
		\centering
		\includegraphics[scale=\sepf]{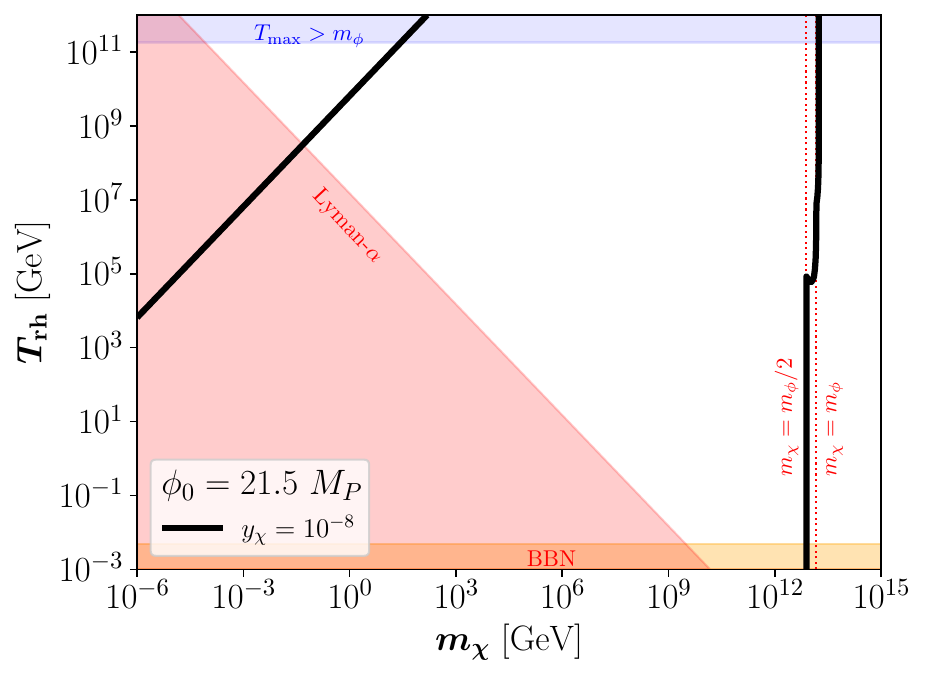}
		\includegraphics[scale=\sepf]{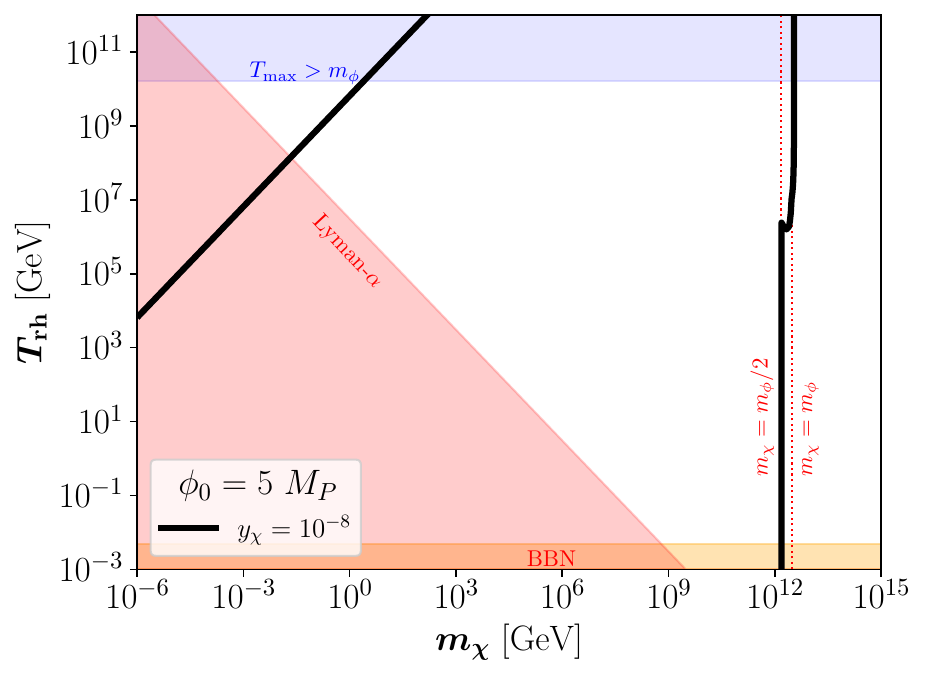}
		\includegraphics[scale=\sepf]{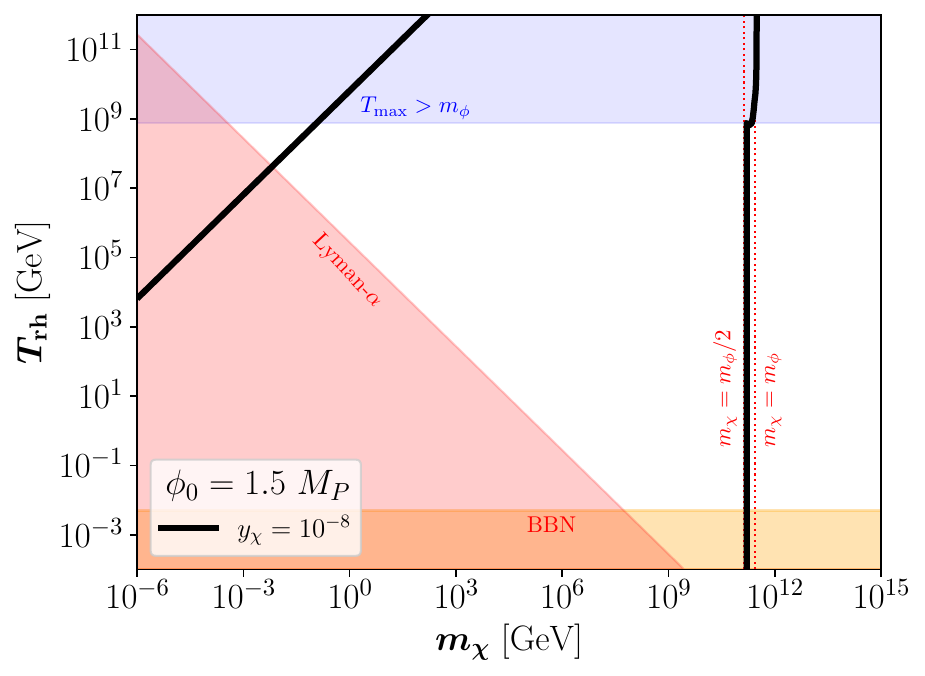}
		\includegraphics[scale=\sepf]{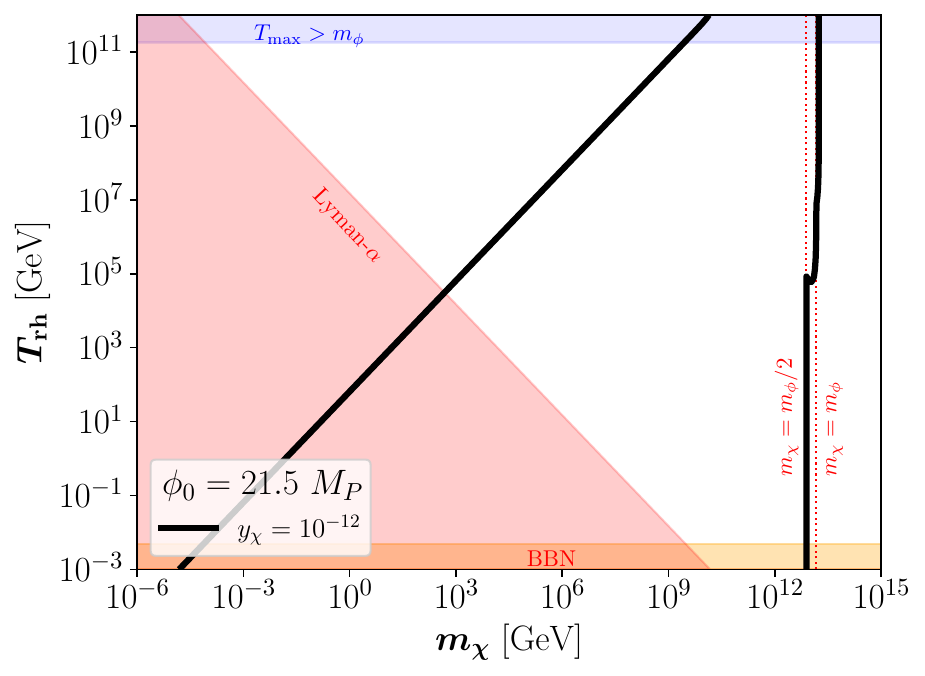}
		\includegraphics[scale=\sepf]{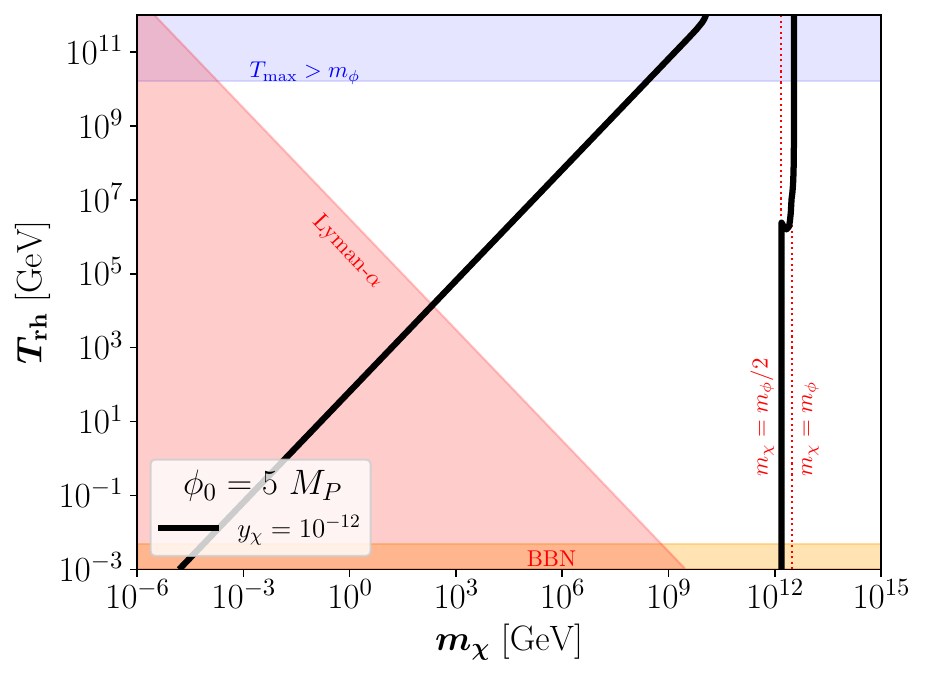}
		\includegraphics[scale=\sepf]{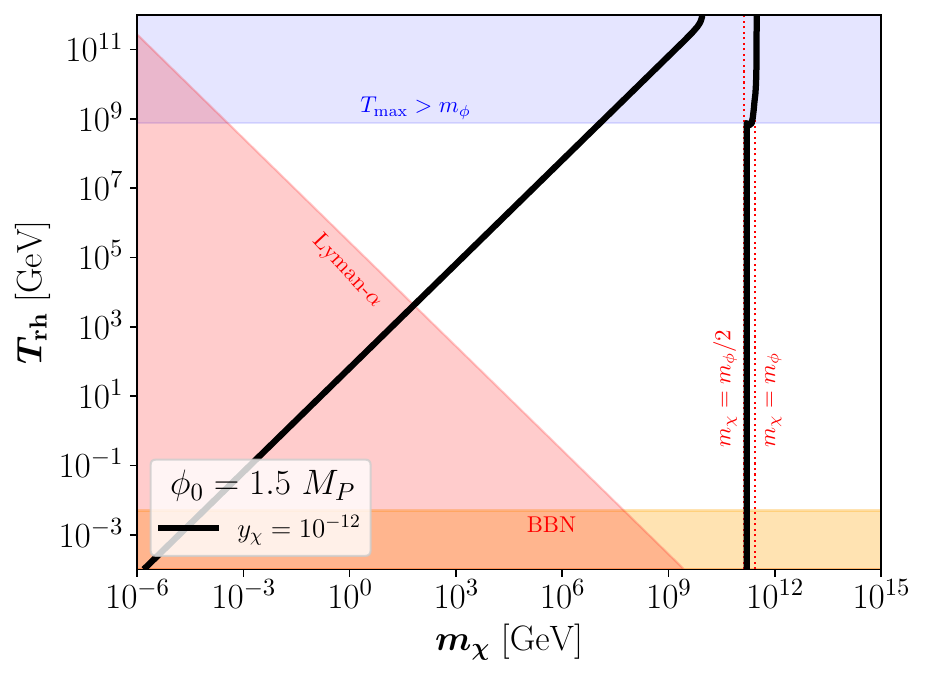}
		\includegraphics[scale=\sepf]{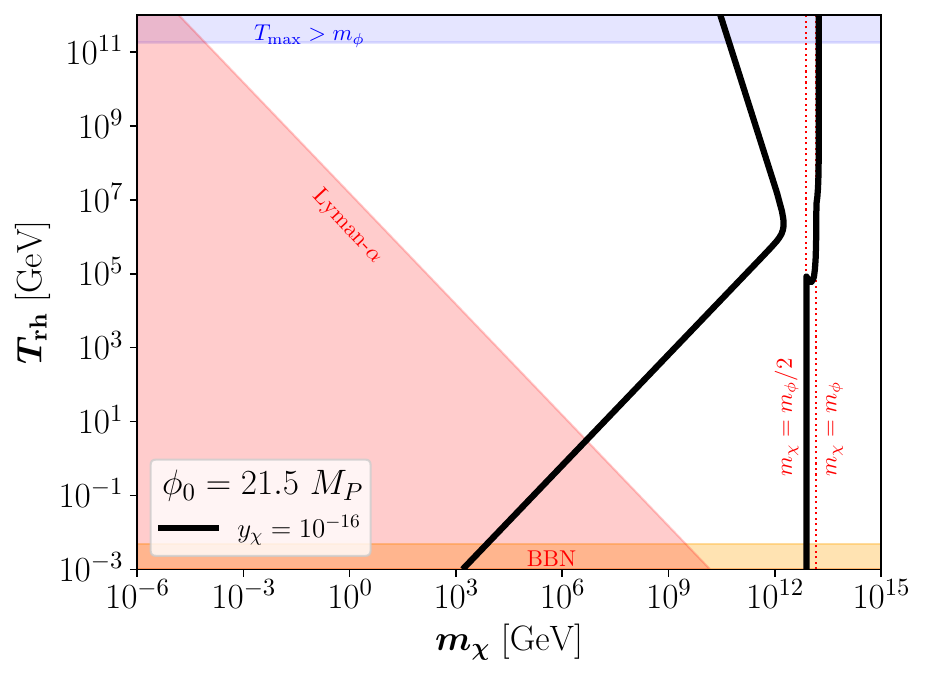}
		\includegraphics[scale=\sepf]{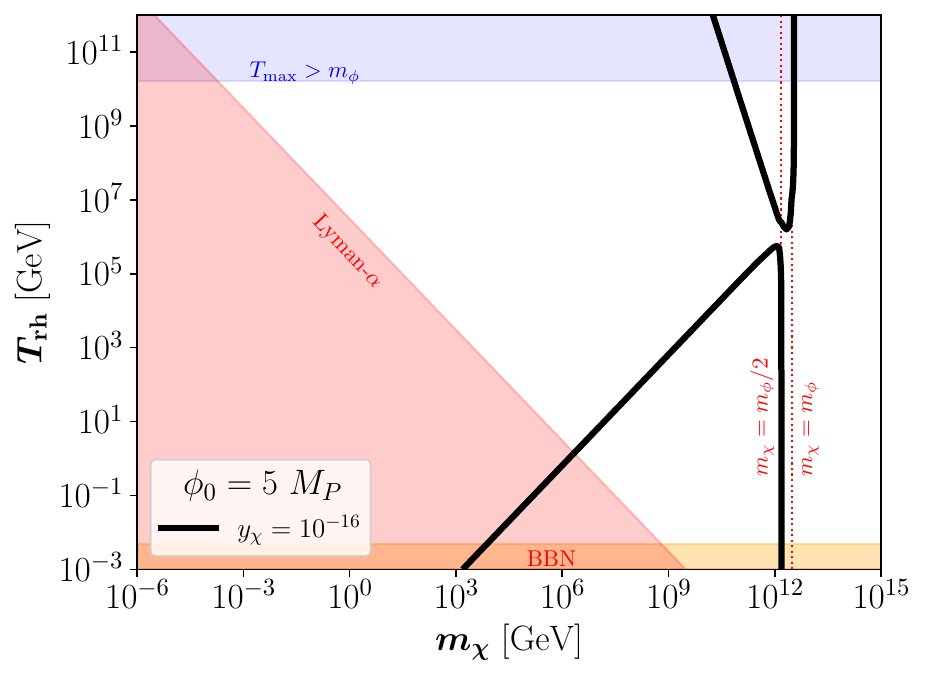}
		\includegraphics[scale=\sepf]{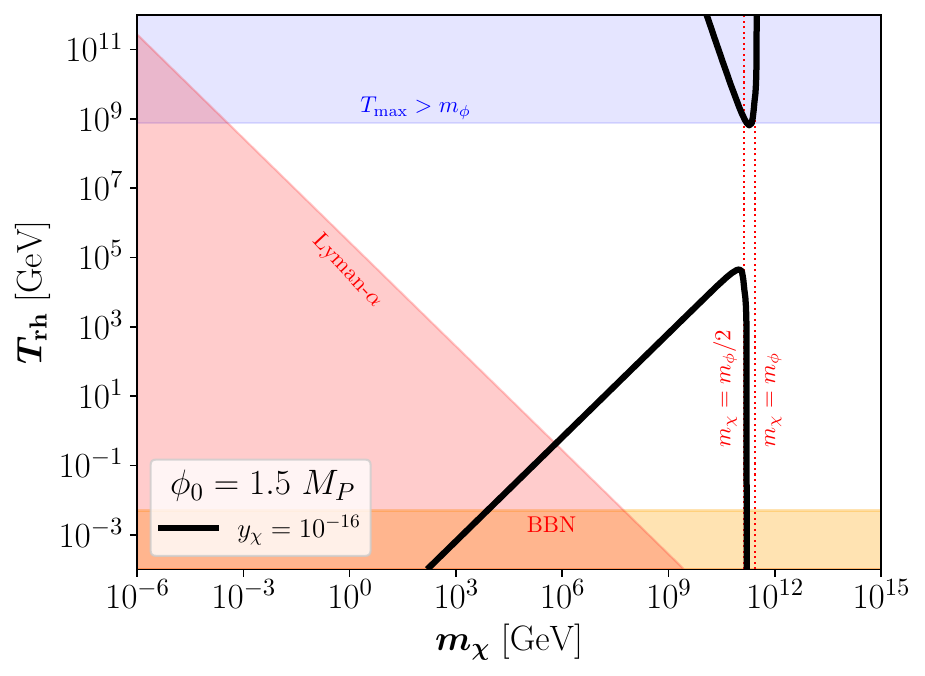}
		\includegraphics[scale=\sepf]{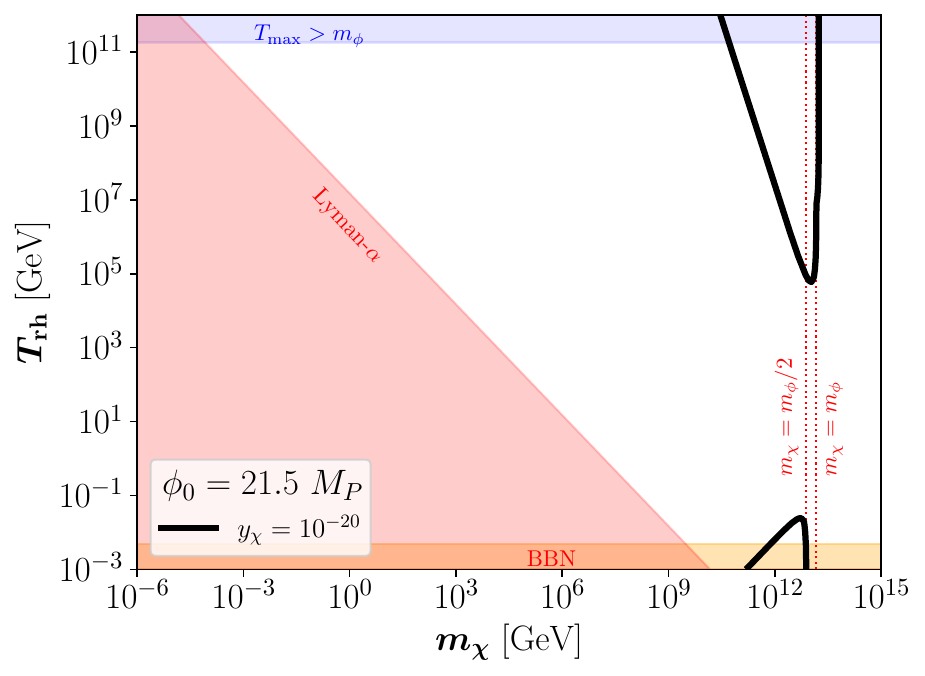}
		\includegraphics[scale=\sepf]{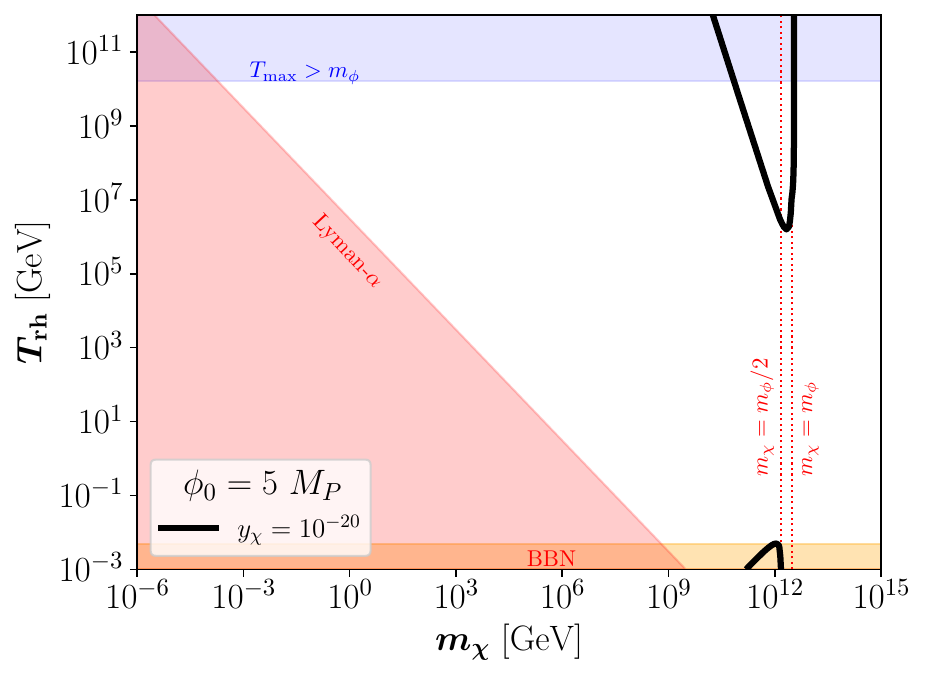}
		\includegraphics[scale=\sepf]{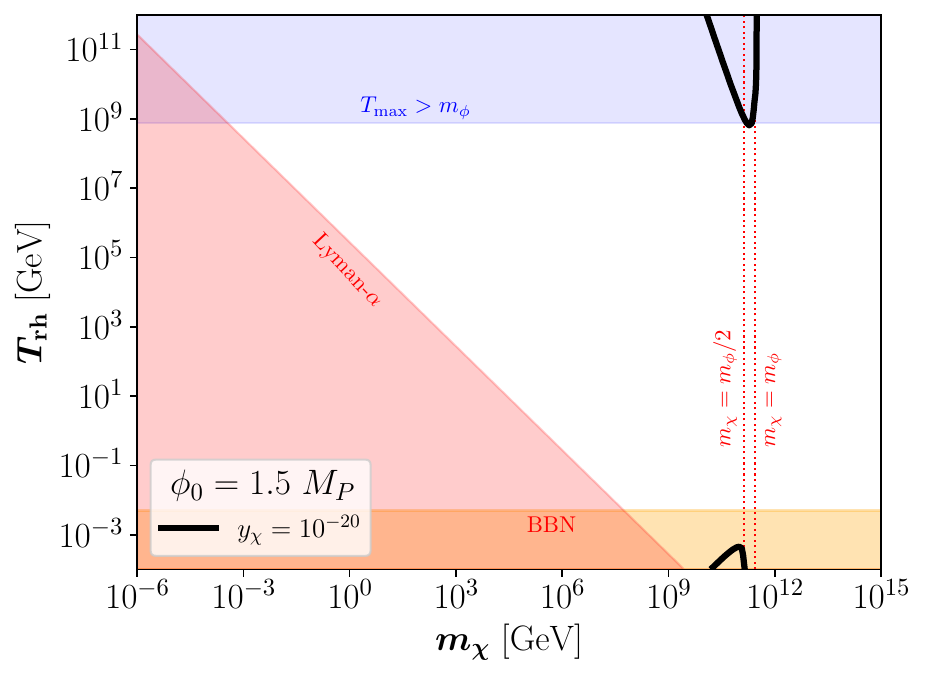}
		\caption{Parameter space (thick black lines) that reproduce the whole observed DM abundance taking into account {\it all} processes discussed in the text, for  different values of the Yukawa coupling $y_\chi$ and $\phi_0 = 21.5~M_P$ (first column), $\phi_0 = 5~M_P$ (second column) and $\phi_0 = 1.5~M_P$ (third column). The colored bands are in tension with BBN (orange), Lyman-$\alpha$ (red) or generate a radiative-unstable potential (blue). The two red dotted vertical lines correspond to $\mdm = m_\phi$ and $\mdm = m_\phi/2$. }
		\label{fig:Decay_vs_Gra}
	\end{figure}

	%%%%%%%%%%%%%%%%%%%%%%%%%%%%%%%%%%%%%%%%%
	In the following, we combine {\it all} previously studied processes and investigate their interplay for different DM masses $m_\chi$ and reheating temperatures $T_{\mathrm{rh}}$. Figure~\ref{fig:Decay_vs_Gra} shows, with thick black lines, the parameter space that reproduces the entire observed DM abundance considering all processes previously discussed. From left to right, we show the results for different values of the Yukawa coupling $y_\chi$ with $\phi_0 = 21.5~M_P$ (first column), $\phi_0 = 5~M_P$ (second column) and $\phi_0 = 1.5~M_P$ (third column). The two red dotted vertical lines correspond to $\mdm = m_\phi$ and $\mdm = m_\phi/2$. 
	%%%%%%%%%%%%%%%%%%%%%%%%%%%%%%%%%
	%%%%%%%%%%%%%%%%%%%%%%%%%%%%%%%%%
	
	We start by considering the largest possible value $\phi_0 = 21.5~M_P$, corresponding to $m_\phi = 1.5\times 10^{13}~\text{GeV}$, shown in the four panels to the left. The upper panel shows the viable parameter space for $y_\chi=10^{-8}$, and the lower panels show how the parameter space changes as the value of $y_\chi$ is lowered. For $y_\chi$ larger than a threshold value, the viable parameter space is continuous in $\Trh$ and discontinuous in $m_\chi$, as seen in the three upper panels. As $y_\chi$ is lowered below a threshold value, the parameter space becomes continuous in $m_\chi$ and discontinuous in $\Trh$. We find numerically that the threshold value is $y_\chi \approx 10^{-17}$, which is in agreement with the analytical threshold value obtained in Eq.~\eqref{eq:ychi_bound_lower}. For $y_\chi\gtrsim 10^{-17}$, the branch to the left corresponds to light DM ($\mdm < m_\phi/2$), which is produced via both inflaton decays and graviton-mediated annihilations of inflatons and SM particles. The branch to the right corresponds to heavier DM masses ($m_\phi/2 \lesssim \mdm < \Tmax$) and is independent of the value of the Yukawa coupling $y_\chi$ for $\Trh \gtrsim 10^5$ GeV, where DM production is dominated by gravitational processes.\footnote{For $\Trh \lesssim 10^5$ GeV, and DM masses very close to the kinematic threshold, DM production is dominated by inflaton decays.} DM is overproduced between the two branches, while it is underproduced in the two complementary regions of parameter space. For $y_\chi \lesssim 10^{-17}$, there is a continuous range of viable DM masses that could account for the whole observed relic abundance, $2\times10^{10}~\text{GeV} \lesssim \mdm < m_\phi$. Here, two branches occur for different values of $\Trh$: If $\Trh \gtrsim 10^5$~GeV, DM is produced by gravitational processes, whereas if $\Trh \lesssim 10^5$~GeV it is generated by the decay of the inflaton. Note that the black lines correspond to the parameter space where the correct relic abundance is reached, while the white region within them corresponds to overproduction. The remaining part leads to underproduction.
	
	As $\phi_0$ decreases, the maximum allowed value for the reheating temperature $\Trh$ decreases, resulting in suppression of the gravitational production channels. Consequently, the interplay between graviton-mediated processes and inflaton decay becomes less prominent and eventually vanishes for smaller values of $\phi_0$, as shown in the second and third columns of Figure~\ref{fig:Decay_vs_Gra}. In practice, we find that for $m_\phi \lesssim 2.8\times10^{11}$~GeV or equivalently $\phi_0 \lesssim 1.5\, M_P$, gravitational channels can no longer account for the DM relic abundance in any part of the parameter space that is consistent with the bound on $\Trh$, as seen in the third column of Fig.~\ref{fig:Decay_vs_Gra}.
	
	%%%%%%%%%%%%%%%%%%%%%%%%%%%%%%%%%%%%%%%%%%
	\section{Conclusions} \label{sec:conclusion}
	%%%%%%%%%%%%%%%%%%%%%%%%%%%%%%%%%%%%%%%%%%
	Polynomial inflation is a simple cosmological framework that is consistent with the cosmic microwave background data and provides testable predictions for the tensor-to-scalar ratio and the running of the spectral index~\cite{Drees:2021wgd, Drees:2022aea}. However, it is desirable that a consistent cosmological model also explains other observations, including the abundance of dark matter (DM). Within the polynomial inflation framework, it was shown in Ref.~\cite{Bernal:2021qrl} that for small field values, the direct decay of the inflaton dominates over all other production mechanisms due to the upper bound on the maximal temperature.  However, the maximum temperature can be orders of magnitude higher in large-field scenarios than in the small-field case. 
	Consequently, DM-production channels that are suppressed in the small-field setup become important in the large-field setup. 
	
	We investigated all relevant DM-production channels, including inflaton decays and inflaton- and graviton-mediated scatterings; see Fig.~\ref{fig:dia_fermion} for the relevant Feynman diagrams. We find that gravitational inflaton annihilations dominate over inflaton decays, for inflaton-DM couplings smaller than the threshold value presented in Eq.~\eqref{eq:ychi_bound_lower}. When considering all production channels simultaneously, the parameter space $(m_\chi,\,\Trh)$ where the whole DM relic abundance is correctly accounted for is generally separated into two disconnected branches. For $y_\chi$ larger than a threshold value, the two branches are separated by a range of values for $m_\chi$ where the correct relic abundance cannot be accounted for. For $y_\chi$ smaller than the threshold, the two branches are separated by a range of values for $\Trh$ where the relic abundance cannot be correctly reproduced, see Fig.~\ref{fig:Decay_vs_Gra} for details. We find that the correct relic abundance can be obtained for DM masses up to $10^{13}~\text{GeV}$, which is two orders of magnitude higher than the heaviest DM mass that can be accommodated in the small field setup.
	
	In summary, the current work offers a comprehensive investigation of DM production after large-field polynomial inflation. The presented formalism can be readily extended to other inflationary setups.
	
	%%%%%%%%%%%%%%%%%%%%%%%%%%%%%%%%%%%%%%%%%%%
	\acknowledgments
	%%%%%%%%%%%%%%%%%%%%%%%%%%%%%%%%%%%%%%%%%%%
	NB received funding from the Spanish FEDER / MCIU-AEI under the grant FPA2017-84543-P. MAM acknowledges support from the DFG Collaborative Research Centre ``Neutrinos and Dark Matter in Astro- and Particle Physics'' (SFB 1258). JH, MAM, and YX acknowledge support from the Cluster of Excellence ``Precision Physics, Fundamental Interactions, and Structure of Matter'' (PRISMA$^+$ EXC 2118/1) funded by the Deutsche Forschungsgemeinschaft (DFG, German Research Foundation) within the German Excellence Strategy (Project No. 390831469).
	
	%%%%%%%%%%%%%%%%%%%%%%%%%%%%%%%%%
	\appendix
	%%%%%%%%%%%%%%%%%%%%%%%%%%%%%%%%%
	%%%%%%%%%%%%%%%%%%%%%%%%%%%%%%%%%
	\section{Gravitational Scattering} \label{eq:appendixA}
	%%%%%%%%%%%%%%%%%%%%%%%%%%%%%%%%%
	In this appendix, we present a detailed derivation of the interaction-rate density for DM production from graviton-mediated inflaton annihilation, corresponding to the first diagram in the second row of Fig.~\ref{fig:dia_fermion}. We label the particle momenta as $\phi(p_1)\,\phi(p_2) \to h_{\mu\nu}(q) \to \chi(p_3)\, \bar{\chi}(p_4)$. The relevant vertices are given by~\cite{Choi:1994ax}
	\begin{equation}
		-\frac{i}{ M_{P}}\left[p_{1 \mu}\,p_{2 \nu}+p_{1 \nu}\, p_{2 \mu}-\eta_{\mu \nu}\left(p_{1} \cdot p_{2}+m_{\phi}^{2}\right)\right]
	\end{equation}
	for $\phi\phi h_{\mu \nu}$, and
	\begin{equation}
		-\frac{i}{4 M_{P}}\left[\left(p_{3}-p_{4}\right)_{\mu} \gamma_{\nu}+\left(p_{3}-p_{4}\right)_{\nu} \gamma_{\mu}-2 \eta_{\mu \nu}\left(\slashed{p}_3 - \slashed{p}_4 -2 m_{\chi}\right)\right]
	\end{equation}
	for $\bar{\chi} \chi h_{\mu \nu}$. The graviton propagator reads
	\begin{equation}
		\Pi^{\mu \nu \rho \sigma}=\frac{1}{2 q^{2}}\left(\eta^{\rho \nu} \eta^{\sigma \mu}+\eta^{\rho \mu} \eta^{\sigma \nu}-\eta^{\rho \sigma} \eta^{\mu \nu}\right).
	\end{equation}
	From these Feynman rules, we readily obtain the squared matrix element
	\begin{equation} 
		\sum_{s_3,s_4} |\mathcal{M}|^2 = \frac{1}{2\, M_P^4} m_\chi^2 (m_\phi^2 - m_\chi^2)\,,
	\end{equation}
	where $s_3, s_4$ denotes the spins of the final states. 
	With the matrix element at hand, we arrive at the final result for the interaction-rate density
	\begin{equation}
		\gamma = \frac{\rho_{\phi}^2}{m_{\phi}^2}\, \frac{ \mdm^2 }{64\pi M_P^4} \left(1-\frac{\mdm^2}{m_{\phi}^2}\right)^{3/2}.
	\end{equation}
	
	%%%%%%%%%%%%%%%%%%%%%%%%%%%%%%%%%
	\bibliographystyle{JHEP}
	\bibliography{biblio}
\end{document}